\documentclass[11pt, twoside, letterpaper]{article}
\usepackage{tikz}
\usepackage{verbatim}
\usepackage{color}
\usepackage{amsmath}
\usepackage{amssymb}
\usepackage{amsthm}
\usepackage{bbm}
\usepackage{algpseudocode}

\newtheorem{theorem}{Theorem}[section]
\newtheorem{lemma}{Lemma}[section]
\newtheorem{definition}{Definition}[section]
\newtheorem{corollary}{Corollary}[section]

\numberwithin{equation}{section}

\newcommand{\eqdef}{\mathrel{:=}}

\usepackage[margin=1in]{geometry}

\title{Funnel theorems for spreading on networks}
\author{Gadi Fibich, Tomer Levin, and Steven Schochet}

\date{\today}

\makeatletter
\let\inserttitle\@title
\makeatother
\usepackage{fancyhdr}
\def\vomega{\mbox{\boldmath $\omega $}}

\begin{document}

		\date{}
	\maketitle

	\vspace{-3em} 
	\begin{center}
	\footnotesize
	 Department of Applied Mathematics, Tel Aviv University \\ fibich@tau.ac.il, levintmr@gmail.com, schochet@tau.ac.il
	\end{center}
	
		\begin{abstract}
	\footnotesize
			We derive novel analytic tools for the
	 Bass and SI models on networks for the spreading of innovations and epidemics on networks.
	 We prove that the correlation between the nonadoption (noninfection) probabilities of $L \ge 2$ disjoint subsets of nodes $\{A_l\}_{l=1}^L$ is non-negative, find the necessary and sufficient condition that determines whether this correlation is positive or zero, and provide an upper bound for its magnitude. Using this result, we prove the funnel theorems, which provide lower and upper bounds for 
	 the difference between the non-adoption probability of a node and the product of its nonadoption probabilities on $L $~modified networks in which the node under consideration is only influenced by incoming edges from~$A_l$ for $l=1, \dots, L$.
	 The funnel theorems can be used, among other things, to explicitly compute the exact expected adoption/infection level on various types of networks, both with and without cycles. 
	\end{abstract}

\section{Introduction}

Mathematical models for the spreading of epidemics have been around for a century~\cite{SIR}. For example, in the Susceptible-Infected (SI) model, the epidemics starts from a few infected individuals 
and progresses as infected individuals transmit it to susceptible ones. 
Mathematical models for the spreading of innovations is a younger problem --- the first model
was introduced in~1969 by Bass~\cite{Bass-69}.  In this model, individuals adopt a new product because
of {external influences} by mass media and internal influences (peer effect, word-of-moth)
by individuals who have already adopted the product.

For many years, the spreading of epidemics and innovations were analyzed using compartmental models, which are typically given by one or several  deterministic ordinary differential equations. Such models implicitly assume that
all the individuals within the population are equally likely to influence each other, i.e., that the underlying social network is a homogeneous complete graph.
In more recent years, 
research on the spreading of epidemics and innovations has gradually shifted 
to network models, in which the adoption/infection event of each individual is stochastic~\cite{Epidemics-on-Network-17,Rand-11}. 
These network models allow for heterogeneity among individuals,  
and for implementing a social network structure, whereby individuals can only influenced by 
their peers.

The SI and Bass models on networks can be solved numerically. They can also be solved analytically using some approximation (mean-field, closure at the level of pairs, etc.).
The focus of this study, however, is on obtaining explicit solutions 
that are exact. To do that, one typically starts from the    
master (Kolmogorov) equations, which are~$2^M-1$ coupled deterministic linear ODEs, where $M$ is the number of nodes. 
To be able to solve this exponentially-large system explicitly, one needs to reduce the number of ODEs significantly. 

At present, there are three analytic approaches for solving the master equations explicitly, without making any approximation. The first is based on {\em utilizing symmetries} of the master equations, in order to reduce the number of equations. This approach was  applied to homogeneous circles, 
and to homogeneous and inhomogeneous complete networks~\cite{OR-10,DCDS-23,Bass-monotone-convergence-23}.   
The second approach is based on the {\em indifference principle}~\cite{Bass-boundary-18}. This analytic tool
simplifies the explicit calculation of adoption probabilities, by replacing the original network with a simpler one.   
The indifference principle has been used to compute the adoption probabilities of nodes on  bounded and unbounded lines, on circles, and on percolated lines~\cite{Bass-percolation-20,Bass-boundary-18}.
The third approach is to identify networks
  on which there is an exact closure at the level of triplets,
  such as undirected graphs with no cycles~\cite{Sharkey-15}.

In this paper we introduce a new approach, is based on the {\em funnel theorems}
 that are derived in Section~\ref{sec:funnel-thms}. 
%
Choose some node~$j$, 
and divide the remaining $M-1$~nodes 
into $L$~subsets of nodes, denoted by $\{A_l\}_{l=1}^L$.
The funnel theorems provide the relation between the nonadoption probability of~$j$ in the original network, and the product of its nonadoption probability on $L$~modified networks in which $j$~can only be directly influenced by edges arriving from~$A_l$, where $l=1, \dots L$. In general, the adoption probabilities on these modified networks are easier to compute,
since the indgree of~$j$ is lower than in the original network.
For example, application of the funnel equality 
on undirected lines reduces this problem to that of directed lines, which is an easier task (Section~\ref{sec:bounded-lines}).

 The funnel relation is an equality if~$j$ is a vertex cut, or more generally if~$j$ is a~{\em funnel node}.
 This is the case, e.g., for any node on an undirected network that does not lie on a cycle. 
 When~$j$ is not a funnel node, however, the funnel relation is a strict inequality. This is the case e.g., for any node on an undirected network that lies on a cycle. 
Indeed, application of the funnel relation leads to a novel strict inequality for the expected adoption/infection level on circular networks
 (Section~\ref{subsec:homogeneous_circles}). 
 
   For the cases where the funnel relation is not an equality, 
 we derive an upper bound for the difference
  between the nonadoption probability of~$j$ in the original network, and the product of its nonadoption probability on the $L$~modified networks. This bound was recently used to show that
   the effect of cycles goes to zero on infinite regular networks and on infinite Erd\H{o}s-Reny\'i networks~\cite{ER-25}.
   Therefore, although these networks have an infinite number of cycles, one can use the funnel equality 
  to compute the exact expected adoption/infection levels on these networks.

To prove the funnel theorems, we first show that 
the correlation between the nonadoption (noninfection) probabilities of $L \ge 2$ disjoint subsets of nodes is non-negative, find the exact condition that determines whether that correlation is positive or zero,
and obtain an upper bound for the magnitude of the correlation.
These results are needed for the proof of the funnel theorems, but are also 
of interest by themselves, and can be used in the analysis of the 
master equations.

To illustrate the power of the funnel theorems, we use them to obtain several new results:
\begin{enumerate}
	\item A novel inequality for circular networks (Lemma~\ref{lem:alpha_q-circle-II}).
	\item An explicit expression for the adoption/infection probability of nodes on bounded lines (Lemma~\ref{lem:fj_twosided_new}).
	
	\item A simple proof that the adoption/infection level on the one-sided line is strictly slower than on two-sided  anisotropic lines
	(Theorem~\ref{thm:f_line^one-sided<f_line_two-sided}).

	\item A  proof that the adoption/infection level on infinite multi-dimensional Cartesian networks is strictly higher than on the infinite line   
	(Lemma~\ref{lem:f_1D_bound}). 
\end{enumerate}

%

  The paper is organized as follows.  Section~\ref{sec:model} describes 
  a unified model for the Bass and SI~models on networks,
  presents the master equations, and introduces the notion of {\em influential nodes}.  In Section~\ref{sec:Sij>SiSj}
  we show that $[S_{\cup_{l=1}^L\Omega_l}]   \ge  \prod_{l=1}^L [S_{\Omega_l}]$,
   find the necessary and sufficient conditions for this inequality to be strict,  obtain an upper bound for this difference, and discuss some applications of these results. Section~\ref{sec:funnel-thms} presents the funnel theorems, which is the main result of this study. Applications of the funnels theorems are given in Section~\ref{sec:applications-funnel}. 
The proofs of all the major theorems are given in Sections~\ref{sec:Sij>SiSj-proof} and~\ref{sec:ptoofs-funnel}.
Section~\ref{sec:Final} concludes with a comparison to related results in the epidemiology literature, and suggestions for future research.

\section{The Bass/SI model on networks}
   \label{sec:model}

The Bass model describes the adoption of new products or innovations within a population. In this framework, all individuals start as non-adopters and can transition to becoming adopters due to two types of influences: external factors, such as exposure to mass media, and internal factors where individuals are influenced by their peers who have already adopted the product.
The SI~model is used to study the spreading of infectious diseases within a population. In this model, some individuals are initially infected (the "patient zero" cases), all subsequent infections occur through internal influences, whereby infected individuals transmit the disease to their susceptible peers, and infected individuals remain contagious indefinitely.
In both models, once an individual becomes an adopter/infected, it remains so at all later times.  In particular, 
she or he remain ``contagious'' forever.  
The difference between the SI~model and the Bass model is the lack of external influences in the former, and the lack of initial adopters in the latter. 

It is convenient to unify these two models into a single model, 
the Bass/SI~model on networks, as follows. 
Consider $M$~individuals, denoted by ${\cal M}:=\{1, \dots, M\}$. 
	We denote by $X_j(t)$ the state of individual~$j$ at time~$t$, so that 
\begin{equation*}
	X_j(t)=\begin{cases}
		1, \qquad {\rm if}\ j\ {\rm is \ adopter/infected \ at\ time}\ t,\\
		0, \qquad {\rm otherwise,}
	\end{cases}
   \qquad j \in \cal M.
\end{equation*} 
	The initial conditions at $t=0$ are stochastic, so that 
\begin{subequations}
	\label{eqs:Bass-SI-models-ME}
	\begin{equation}
		\label{eq:general_initial}
		X_j(0)=	X_j^0 \in \{0,1\}, \qquad j\in {\cal M},
	\end{equation}
	where
\begin{equation}
	\mathbb{P}(X_j^0=1) =I_j^0, \quad 
	\mathbb{P}(X_j^0=0) =1-I_j^0,\quad I_j^0 \in [0, 1],  \qquad 
	j \in \cal M,
\end{equation}
and 
\begin{equation}
	\label{eq:p:initial_cond_uncor-two_sided_line}
	\mbox{the random variables $\{X_j^0 \}_{j \in \cal M}$ are independent}.
\end{equation} 
Deterministic initial conditions are a special case where
$I_j^0 \in \{0,1\}$. 

So long that $j$ is not a nonadopter/susceptible, its adoption/infection rate at time~$t$ is
	\begin{equation}
		\label{eq:lambda_j(t)-Bass-model-heterogeneous-tools}
		\lambda_j(t) = p_j+\sum\limits_{k\in {\cal M}} q_{k,j} X_{k}(t),
		\qquad j \in {\cal M}.
	\end{equation}
	Here, $p_j$ is the rate of external influences on~$j$, and~$q_{k,j}$ is the rate of 
internal influences ({\em peer effects}) by~$k$ on~$j$, provided that $k$ is already an adopter/infected. Once~$j$ adopts the product/becomes infected, it remains so at all later times.\,\footnote{i.e., the only admissible transition is 
$X_j=0 \to X_j=1$}
	Hence, as $ \Delta t \to 0$,
	\begin{equation}
		\label{eq:general_model}
		\mathbb{P} (X_j(t+\Delta  t )=1  \mid   {\bf X}(t))=
		\begin{cases}
			\lambda_j(t) \, \Delta t , &  {\rm if}\ X_j(t)=0,
			\\
			1,\hfill & {\rm if}\ X_j(t)=1,
		\end{cases}
		\qquad 	j \in {\cal M},
	\end{equation}
	where ${\bf X}(t) := \{X_j(t)\}_{j \in \cal M}$ 
	is the state of the network at time~$t$,
	and
	\begin{equation}
		\label{eq:Bass-SI-models-ME-independent}
		\mbox{the random variables $\{X_j(t+\Delta  t )  \mid   {\bf X}(t) \}_{j \in \cal M}$ are independent}.
	\end{equation}  
	%
	%
We assume that all the nodes have a positive probability to 
be initially nonadopters/susceptible, and that the external and internal influence rates are non-negative, i.e.,   
\begin{equation}
\label{eq:parameters-assumtions-A}
	0 \le I_j^0<1, \quad  p_j \ge 0, \quad q_{k,j} \ge 0, \qquad k,j \in \cal M.
\end{equation}
In addition, any node can adopt externally, either at $t=0$ or at $t>0$, i.e., 
\begin{equation}
\label{eq:parameters-assumtions-B}
 	I_j^0>0   \quad \text{or} \quad p_j > 0, \qquad  j \in \cal M. 
\end{equation}
\end{subequations} 
In the Bass model there are no adopters when the product is first introduced into the market, and so $I_j^0 \equiv 0$.
In the SI~model there are only internal influences for $t>0$, and so $p_j= 0$.

 The internal adoption rates $\{q_{k,j}\}$ induce a {\em directed weighted graph} on the nodes~$\cal M$, 
so that node~$j$ has weight~$p_j$ and initial condition~$I_j^0$, the directed edge $k \to j$ exists if and only if~$q_{k,j}>0$,
and its weight is given by~$q_{k,j}$.  
We denote the network that corresponds to~\eqref{eqs:Bass-SI-models-ME} by
${\cal N} = {\cal N}({\cal M},\{p_j\},\{q_{k,j}\},\{I_j^0\})$.

\subsection{Master equations}


The starting point of most of the analytic theory of the  Bass/{\rm SI} model~\eqref{eqs:Bass-SI-models-ME} on networks
are the master equations. 
Let $ \emptyset \not= \Omega \subset  {\cal M}$ be a nontrivial subset of the nodes,  and let
$
\Omega^{\rm c}:={\cal M} \setminus \Omega
$ denote the complementary set.
Let 
\begin{equation}
	\label{eq:Xdef}
	X_\Omega(t):= \max_{k \in \Omega}X_k(t).
\end{equation}
Thus, $	X_\Omega = 0$ if none of the nodes in~$\Omega$ are adopters at time $t$,
and $X_\Omega = 1$ if at least one of the nodes in $\Omega$ is an adopter.
 let 
\begin{equation}
	\label{eq:def-S_Omega-ME}
	S_{\Omega}(t):=\{X_{\Omega}(t)=0 \},
	\qquad  [S_{\Omega}](t)
	:= \mathbb{P}(S_{\Omega}(t)),
\end{equation}
denote the event that all nodes in~$\Omega$ are nonadopters/susceptible at time~$t$, and
the probability of this event, respectively.
	To simplify the presentation, we introduce the notation
	\begin{equation*}
		S_{\Omega_1, \cdots, \Omega_L}:=S_{\cup_{l=1}^L \Omega_l}, \qquad
		\Omega_1, \cdots, \Omega_L \subset \cal M.
	\end{equation*}
	Thus, for example, $	S_{\Omega,k}:=S_{\Omega \cup \{k\}}$ and
	$S_{m_1,m_2, m_3}:= S_{\{m_1\} \cup \{m_2\} \cup  \{ m_3\}}$. 
	We also denote the sum of the external influences on the nodes in~$\Omega$
	and the sum of the internal influences by node $k \in \Omega^c$ on the nodes in~$\Omega$ by 
	$$
	p_\Omega:=\sum_{m  \in \Omega} p_{m},
	\qquad 
	q_{k,\Omega}:=\sum_{m \in \Omega} q_{k,m},
	$$
	respectively.   We then have
	\begin{theorem}[\cite{MOR-23}]
		\label{thm:master-eqs-general}
		The master equations for the Bass/{\rm SI} model~\eqref{eqs:Bass-SI-models-ME} are
		\begin{subequations}
			\label{eqs:master-eqs-general}
			\begin{equation}
				\label{eq:master-eqs-general}
				\frac{d [S_{\Omega}]}{dt}  = 
				-\bigg( p_\Omega
				+\sum_{k \in \Omega^{\rm c}} q_{k,\Omega}\bigg)[S_{\Omega}] 
				+\sum_{k \in \Omega^{\rm c}}q_{k,\Omega}\,  [S_{\Omega,k}],
			\end{equation}
			subject to the initial conditions
			\begin{equation}
				\label{eq:master-eqs-genera-icl}
				[S_{\Omega}](0)=[S_{\Omega}^0], \qquad [S_{\Omega}^0]:= \prod_{m \in \Omega} (1-I_m^0), 
			\end{equation}
			for all $\emptyset\not=\Omega \subset {\cal M}$.
		\end{subequations}
	\end{theorem}
The number of master equations~\eqref{eqs:master-eqs-general}  is $2^M-1$,
which is the number of nontrivial subsets of $M$~modes.

\subsection{Influential edges and nodes}

   Let us recall
   \begin{definition}[influential edge~\cite{Bass-boundary-18}]
   	\label{def:influential-edge}
   	Consider the Bass/{\rm SI} model~\eqref{eqs:Bass-SI-models-ME}. 
   	Let $\Omega \subset \cal M$.
   	A directed edge  $k \to m$  is said to be ``influential to $\Omega$" 
   if~$k\in \Omega^{\emph{\text{c}}}$, and if 
   	 either $m \in \Omega$ or there is a path from~$m$ to~$\Omega$ which does not go through the node~$k$.
     	Any edge which is not ``influential to~$\Omega$'' is called   
   	``non-influential to $\Omega$".
   \end{definition}
We then have
\begin{theorem}[Indifference principle~\cite{Bass-boundary-18}]
	\label{thm:Indifference}
	Consider the Bass/{\rm SI}  model~\eqref{eqs:Bass-SI-models-ME}. 
	Let $\emptyset \not=\Omega \subset {\cal M}$.
	Then $[S_{\rm \Omega}]$
	remains unchanged if we delete or add edges that are non-influential to~$\Omega$.
\end{theorem}

We also introduce a new notion of influential node:
\begin {definition} [influential node] 
\label{def:influential_nodes} 
Let $\emptyset \not= \Omega \subset {\cal M}$. 
We say that ``node~$m$ is influential to~$\Omega$'' if 
 $m \in \Omega$, or if 
 $m \in \Omega^{\rm c}$ and there is path
	from~$m$ to~$\Omega$.
\end{definition}

The following result is immediate:
\begin{lemma}
	\label{lem:paths-influential-node}
	Let $\emptyset \not=\Omega_1, \Omega_2 \subset {\cal M}$ such that $\Omega_1 \cap \Omega_2 = \emptyset$. Then 
there exists a node in~${\cal M}$ which is influential to both~$\Omega_1$ and~$\Omega_2$
	if and only if  at least one of the following conditions hold:
	\begin{enumerate}
		\item There exists a path  from~$\Omega_1$ to~$\Omega_2$ or   from~$\Omega_2$ to~$\Omega_1$.
		\item There exists a node~$m \notin \Omega_1 \cup \Omega_2$ from which there exist a path to~$\Omega_1$ and a path to~$\Omega_2$.
	\end{enumerate}
\end{lemma}

\begin{corollary}
	\label{cor:paths-influential-node}
   	Let $\emptyset \not=\Omega_1, \Omega_2 \subset {\cal M}$ such that $\Omega_1 \cap \Omega_2 = \emptyset$. 
	Let the network be undirected. Then there exists a node which is influential to both~$\Omega_1$ and~$\Omega_2$
	if and only if there exists a path between~$\Omega_1$ and~$\Omega_2$.
	
\end{corollary}

\section{Lower and upper bounds for $[S_{\cup_{l=1}^L\Omega_l}] - \prod_{l=1}^L [S_{\Omega_l}]$}
  \label{sec:Sij>SiSj}

It is reasonable to assume that the nonadoption probabilities of $L \ge 2$ disjoint sets of nodes are uncorrelated when the sets are ``disconnected'', and positively-correlated otherwise. 
The rigorous formulation and proof of this statement is as follows:
\begin{theorem}
	\label{thm:[Si,Sj]>[Si][Sj]-K}
	Consider the Bass/{\rm SI}  model~\eqref{eqs:Bass-SI-models-ME}. 
	Let $\Omega_1, \cdots, \Omega_L \subset {\cal M}$, such that $\Omega_l \cap \Omega_{\widetilde{l}}= \emptyset$ if $l \not = \widetilde{l}$. Then 
	\begin{equation}
		[S_{\cup_{l=1}^L\Omega_l}]  \ge \prod_{l=1}^L [S_{\Omega_l}], \qquad t  \ge  0.
	\end{equation}
	In addition, 
	\begin{enumerate}
		\item If  there exist $l,  \widetilde{l} \in \{1, \dots, L\}$  where
		$l \not = \widetilde{l}$, and a node in~${\cal M}$ which is influential to both~$\Omega_l$ and~$\Omega_{\widetilde{l}}$,
		then 
		\begin{equation}
			\label{eq:[S_Omega_ij]>[S_Omega_i][S_Omega_j]-K}
			[S_{\cup_{l=1}^L\Omega_l}]  > \prod_{l=1}^L [S_{\Omega_l}], \qquad t > 0.
		\end{equation}
		\item If, however, for every $l,  \widetilde{l} \in \{1, \dots, L\}$  where
		$l \not = \widetilde{l}$, 
		there is no node in~${\cal M}$ which is influential to both~$\Omega_l$ and~$\Omega_{\widetilde{l}}$, then 
		\begin{equation}
			\label{eq:[S_Omega_ij]=[S_Omega_i][S_Omega_j]-K}
			[S_{\cup_{l=1}^L\Omega_l}]  = \prod_{l=1}^L [S_{\Omega_l}], \qquad t \ge 0.
		\end{equation}
	\end{enumerate}
\end{theorem}
\begin{proof}
	See Section~\ref{sec:[Si,Sj]>[Si][Sj]-K}.
\end{proof}

We can also derive an upper bound 
for $[S_{\cup_{l=1}^L\Omega_l}] - \prod_{l=1}^L [S_{\Omega_l}]$, 
by utilizing the 
spatio-temporal estimates for the correlation between the adoption probabilities of two nodes, 
which were recently derived in~\cite{OR-24}.  
To simplify the presentation, we assume that 
\begin{enumerate}
	\item All the nodes have the same weight and initial condition, i.e., 
\begin{subequations}
	\label{eqs:assumptions-upper-bound-SiSj}
	\begin{equation}
		p_k \equiv p, \quad  I_k^0 \equiv I^0, \qquad k \in \cal M.
	\end{equation}  
	\item The network is undirected, and that 
	all the edges have same weight, i.e.,
	\begin{equation}
		q_{k,j} = q_{j,k}  \in \{0,q\} , \qquad k,j\in \cal M.
	\end{equation}
   \item The parameters satisfy, see~\eqref{eq:parameters-assumtions-A}
	and~\eqref{eq:parameters-assumtions-B}, 
	\begin{equation}
		q>0, \quad  p \ge 0,  \quad 0 \le I^0 <1,  \quad \text{such that} \quad
		p>0   \quad \text{or} \quad I^0>0.
	\end{equation}
\end{subequations}
\end{enumerate}

We then have 
\begin{theorem}
	\label{thm:[S_Omega]-prod[S_i]-N-paths}
	Consider the Bass/{\rm SI} model~\eqref{eqs:Bass-SI-models-ME} 
	on an undirected network, such that 
	\eqref{eqs:assumptions-upper-bound-SiSj} holds. 
	Let $\Omega_1, \cdots, \Omega_L \subset {\cal M}$, such that
	$\Omega_l \cap \Omega_{\widetilde{l}}  = \emptyset$ for any $l \not=\widetilde{l}$.  
	Denote by~$\{\Gamma_n\}_{n=1}^{N_L}$ the $N_L$~distinct simple paths that connect between pairs of sets in $\{\Omega_1, \dots, \Omega_L \}$,
	such that the interior nodes of~$\{\Gamma_n\}_{n=1}^{N_L}$ are in~${\cal M} \setminus \bigcup_{l=1}^L \Omega_l$.
	Let $N_L \ge 1$. 
	Then
	\begin{equation}
		\label{eq:[S_Omega]-prod[S_i]-N-paths}
		0<[S_{\cup_{l=1}^L\Omega_l}]-\prod_{l=1}^L [S_{\Omega_l}] 
		< 
		\sum_{n=1}^{N_L} 	E(t;K_n), \qquad t > 0  ,	
	\end{equation}
	where $K_n$ is the number of nodes of the path~$\Gamma_n$,\,\footnote{including the two boundary nodes in $ \cup_{l=1}^L \Omega_l$}
	and
	\begin{subequations}
		\label{eqs:[S_i,j]-[S_i][S_j]-N-paths-global-in-time-E}
		\begin{equation}
			\label{eq:[S_i,j]-[S_i][S_j]-N-paths-global-in-time-E-temporal}
			E(t;K) \le 2 (1-I^0) e^{-\left(p+ q \right)t} \bigg(\frac{e q t}{\left \lfloor{\frac{K+1}{2}}\right \rfloor}\bigg)^{\left \lfloor{\frac{K+1}{2}}\right \rfloor}, 
			\qquad    0 < t  < \frac1q  \bigg \lfloor{\frac{K+1}{2}} \bigg \rfloor.
	\end{equation}
	We also have the global bound 
	\begin{equation}
		\label{eq:[S_i,j]-[S_i][S_j]-N-paths-global-in-time-E-global}
		E(t;K) \le 2(1-I^0) \left(\frac{q}{p+q} \right)^{\left \lfloor{\frac{K+1}{2}}\right \rfloor},
		\qquad  t \ge 0 . 
	\end{equation}
\end{subequations}

\end{theorem}
\begin{proof}
	See Section~\ref{sec:[S_Omega]-prod[S_i]-N-paths}.
\end{proof}

\subsection{Applications}
   \label{sec:applications-SiSj}

   In Section~\ref{sec:funnel-thms} we will use the lower and upper bounds for $[S_{\Omega_1, \dots, \Omega_L}]-\prod_{l=1}^L [S_{\Omega_l}]$ that were derived  in Theorems~\ref{thm:[Si,Sj]>[Si][Sj]-K} and~\ref{thm:[S_Omega]-prod[S_i]-N-paths},
   to prove the funnel theorems. 
   The funnel theorems have numerous applications, see Section~\ref{sec:applications-funnel}. In addition, the bounds  for $[S_{\Omega_1, \dots, \Omega_L}]-\prod_{l=1}^L [S_{\Omega_l}]$  can be used in the analysis of the master equations~\eqref{eqs:master-eqs-general}.\footnote{ 
    For example, in the proof of the universal upper bound for spreading on networks in~\cite{Chaos-24}, 
    we wrote the  master equation~\eqref{thm:master-eqs-general} for~$\Omega= j$ as 
     	$[S_j] = e^{-(p_j+q_j)t}+\int_0^te^{-(p_j+q_j)\left(t-\tau\right)}
    	\sum_{k\in \cal M} q_{k,j} \, [S_{k,j}](\tau)\, d\tau$,  
    where $ q_j:=\sum_{k\in \cal M} q_{k,j}$. Then we used Theorem~\ref{thm:[Si,Sj]>[Si][Sj]-K} to bound~$[S_j]$ from below, i.e., 
    $[S_j] \geq e^{-(p_j+q_j)t}+\int_0^te^{-(p_j+q_j)(t-\tau)}\sum\limits_{k\in \cal M}q_{k,j} \, [S_j](\tau) \, [S_k](\tau) \, d\tau $.
}

\section{Funnel  theorems}
\label{sec:funnel-thms}

The funnel theorems make use of the following partition of the nodes:
\begin{definition}[partition of nodes]
	Let $j \in {\cal M}$ and $\emptyset \not= A_l\subset {\cal M} \setminus \{j\}$ for $l=1, \dots,L$. 
	We say that $\{A_1, \cdots,  A_L,\{j\} \}$ is a partition of~${\cal M}$, if $A_1 \cup \dots \cup  A_L \cup \{j\} = {\cal M}$, and  if $\{A_l\}_{l=1}^L$ are mutually disjoint. 
\end{definition}
	Let us introduce the network ${\cal N}^{A,p_j}$, on which~$j$ can adopt
due to internal influences of direct edges from~$A$ and external influences: 
\begin{definition} [Network~${\cal N}^{A,p_j}$]
	\label{def:more-networks}
	Consider the Bass/{\rm SI}  model~\eqref{eqs:Bass-SI-models-ME} on a network~$
	{\cal N}$.  
	Let $j \in \cal M$ and $A \subset {\cal M}$.
	The network		${\cal N}^{A,p_j}$  is obtained from~${\cal N}$ by removing all internal influences on~$j$ from edges 
		that do not emanate from~$A$, i.e., by setting~$q_{k,j} := 0$ for~$k\in {\cal M} \setminus A$.
	\end{definition}
On the network ${\cal N}^{p_j}$, node~$j$ can only adopt
due to external influences: 		
\begin{definition} [Network~${\cal N}^{p_j}$]
	\label{def:networks-N^p_j}
	Consider the Bass/{\rm SI}  model~\eqref{eqs:Bass-SI-models-ME} on 
	network~${\cal N}$.  
		Let $j \in \cal M$.
The network ${\cal N}^{p_j}$ is obtained from~${\cal N}$  by removing all internal influences on~$j$, i.e.,
		by setting~$q_{k,j} := 0$ for~$k\in \cal M$.
\end{definition}
We denote the state and nonadoption probability of~$j$ in these networks
by
$$
X^{U}_j(t):=X^{N^U}_j(t), \quad 
[S^{U}_j] :=\mathbb{P}(X^{U}_j(t)=0) 
\qquad U \in \{  \{A, p_j\}, p_j\},
$$
respectively. We then have the following inequality:
\begin{theorem}
	\label{thm:funnel_node_inequality-frac}
	Consider the Bass/{\rm SI}  model~\eqref{eqs:Bass-SI-models-ME}.
	Let $j \in \cal M$, and let $\{A_1, \dots, A_L,\{j\}\}$ be a partition of $ {\cal M}$.
	Then 
	\begin{equation}
		\label{eq:funnel_cor}
		[S_{{j}}]  \ge
		\frac{\prod_{l=1}^L [S^{A_l, p_j}_j]}{([S^{p_j}_j])^{L-1}},  \qquad t \ge 0,    \qquad  \mbox{\bf (funnel  inequality)}
	\end{equation}
	where 	
	\begin{equation}
		\label{eq:funnel_p_j}
		[S^{p_j}_j] = (1-I^{0}_j) \, e^{-p_j t}.
	\end{equation}
\end{theorem}
\begin{proof}
	See Section~\ref{sec:ptoofs-funnel}.
\end{proof}
	
In order to determine the conditions under which the funnel  inequality becomes an equality, we  introduce 
\begin{definition} [funnel node] 
	\label{def:funnel} 
	Let $\{A_1, \dots, A_L,\{j\}\}$ be a partition of~${\cal M}$.  A node~$j$ is called a ``funnel node of~$\{A_l\}_{l=1}^L$ in network~${\cal N}$'', if there is no node in~${\cal M} \setminus \{j\}$ which is influential to~$j$ both in~${\cal N}^{A_l}$ and in~${\cal N}^{A_{\widetilde{l}}}$ for some $l \not = \widetilde{l}$.
\end{definition}

Recall also the following terminology from graph theory:
\begin{definition} [vertex  cut]
	\label{def:vertex-cut} 
	Let $\{A_1, \dots, A_L,\{j\}\}$ be a partition of~${\cal M}$.  A node~$j$ is called a  ``vertex  cut 
	between~$\{A_l\}_{l=1}^L$,'' if removing~$j$ from the network makes the sets~$\{A_l\}_{l=1}^L$ disconnected from each other.
\end{definition}

Any node which is a vertex cut is also a funnel node:
\begin{lemma}
	\label{lem:j-is-funnel}
	Let $\{A_1, \dots, A_L,\{j\}\}$ be a partition of~${\cal M}$.
	If node~$j$ is  a  vertex cut between~$\{A_l\}_{l=1}^L$,
	then $j$ is a  funnel node.
\end{lemma}
\begin{proof}
	Let~$j$ be  a  vertex cut between~$\{A_l\}_{l=1}^L$.
	If node $m \in A_l$ is influential to~$j$, then~$m$ cannot be 
	influential to~$j$ in~${\cal N}^{A_{\widetilde{l}}}$, since in~${\cal N}^{A_{\widetilde{l}}}$
	we removed all the edges from $A_l$ to~$j$, and there is no sequence of edges (influential or not) from~$m$ to~$A_{\widetilde{l}}$. 
\end{proof}


The converse statement, however, is not true, i.e., 
there are networks in which $j$ is a funnel node, yet the sets~$A_l$ and~$A_{\widetilde{l}}$ are directly connected. For example, this is the case if all edges
between~$A_l$ and~$A_{\widetilde{l}}$ are non-influential to~$j$.  Moreover, even 
if there are two nodes $m \in A_l$ and~$\widetilde{m} \in A_{\widetilde{l}}$ that are connected by an influential edge
$ m \to \widetilde{m} $, $j$~may still be a funnel node, provided that there is no influential edge that emanates from node~$m$ in network~${\cal N}^{A_l}$ (if there exists an influential edge emanating 
from~$m$, then $j$ is not a funnel node).


\begin{theorem}
	\label{thm:funnel_node_inequality-frac-B}
	Assume the conditions of Theorem~{\rm \ref{thm:funnel_node_inequality-frac}}.  
	\begin{itemize}
		\item 
		
		If  $j$ is a funnel node of~$\{A_l\}_{l=1}^L$,  then 
		\begin{equation}
			\label{eq:funnel_equality-A-p}
			[S_{{j}}]  =
			\frac{\prod_{l=1}^L [S^{A_l, p_j}_j]}{([S^{p_j}_j])^{L-1}},  \qquad t \ge 0.  \qquad  \mbox{\bf (funnel  equality)}
		\end{equation}
		
		\item 
		If, however, $j$ is not a funnel node of~$\{A_l\}_{l=1}^L$, then
		\begin{equation}
			\label{eq:funnel_strict_inequality-A-p}
			[S_{{j}}]  >
			\frac{\prod_{l=1}^L [S^{A_l, p_j}_j]}{([S^{p_j}_j])^{L-1}},  \qquad t > 0.  \qquad  \mbox{\bf (strict funnel inequality)}
		\end{equation}
	\end{itemize}
\end{theorem}
\begin{proof}
	See Section~\ref{sec:ptoofs-funnel}.
\end{proof}

\begin{corollary}
	\label{cor:1path-from-m-to-j-funnel}
	Assume the conditions of Theorem~{\rm \ref{thm:funnel_node_inequality-frac}}.  
	Let $ j \in \cal M$.
	If for any $m \in {\cal M} \setminus \{j\}$, there is at most one finite path from~$m$ to~$j$, then
	the funnel equality~\eqref{eq:funnel_equality-A-p} holds.    
\end{corollary}
\begin{proof}
	If there exists node $m$ which is influential to~$j$ in~${\cal N}^{A_l}$  and in~${\cal N}^{A_{\widetilde{l}}}$, then 
	there are two different paths leading from~$m$ to~$j$. Therefore, there is no such node~$m$. Hence, $j$ is a funnel node of~$\{A_1\}_{l=1}^L$,
	and so the result follows from Theorem~\ref{thm:funnel_node_inequality-frac-B}.
\end{proof}

\begin{corollary}
	\label{cor:no-loops-funnel}
	Assume the conditions of Theorem~{\rm \ref{thm:funnel_node_inequality-frac}}. 
	If the network is undirected and contains no cycles,
	the funnel equality~\eqref{eq:funnel_equality-A-p} holds for all $j \in \cal M$.    
\end{corollary}
\begin{proof}
	This follows from Corollary~\ref{cor:1path-from-m-to-j-funnel}. 
\end{proof}


Theorems~\ref{thm:funnel_node_inequality-frac} and~\ref{thm:funnel_node_inequality-frac-B} provide 
a lower bound for $[S_{{j}}]  -\frac{\prod_{l=1}^L [S^{A_l, p_j}_j]}{([S^{p_j}_j])^{L-1}}$. We can also  derive an upper bound for this difference:
\begin{theorem}
	\label{thm:funnel-upper-bound-frac}
	Consider the Bass/{\rm SI}  model~\eqref{eqs:Bass-SI-models-ME}
	on an undirected network,
	such that \eqref{eqs:assumptions-upper-bound-SiSj}~holds. 
	Let $\{A_1, \dots,  A_L,\{j\}\}$ be a partition of~$ {\cal M}$.
	Assume that there are $N_j \ge 1$~cycles $\{C_n\}_{n=1}^{N_j}$ in which
	$j$~is connected to some~$A_{l}$ on one side and to some~$A_{\widetilde{l}}$ on the other side, where $l\not=\widetilde{l}$.\,\footnote{$A_{l}$ and $A_{\widetilde{l}}$ may be different for each cycle.}
	%
	Then 
	\begin{equation}
		\label{eq:funnel-upper-bound-1cycle}
		0<	[S_{{j}}]  -\frac{\prod_{l=1}^L [S^{A_l, p_j}_j]}{([S^{p_j}_j])^{L-1}} 
		<
		[S_j^{p_j}] \, \sum_{n=1}^{N_j} E(t;K_n+1), \qquad t>0, 
	\end{equation}
	where 
	$K_n$ is the number of nodes of~$C_n$,
	and~$E(t;K_n)$ satisfies the bounds \eqref{eq:[S_i,j]-[S_i][S_j]-N-paths-global-in-time-E-temporal} and~\eqref{eq:[S_i,j]-[S_i][S_j]-N-paths-global-in-time-E-global}.
\end{theorem}
\begin{proof}
	See Section~\ref{sec:ptoofs-funnel}.
\end{proof}

\section{Applications of the funnel theorems}
  \label{sec:applications-funnel}
  
We now present several applications of the funnel theorems. 

\subsection{Explicit solutions of the master equations}

The funnel theorems enable us to solve the master equations explicitly for various network types. Briefly, this is because the indegree of node~$j$
on the modified networks $\{{\cal N}^{A_l,p_j}\}_{l=1}^L$ is lower than on the original network. 
For example,  we can use the funnel 
equality to obtain an explicit solution for the adoption/infection probability
of nodes on lines (Lemma~\ref{lem:1D2SWithP_R-hom}) and on a star-shaped network
(Lemma~\ref{lem:intersection}).  More generally, 
 the funnel equality can be applied to any undirected
 network that has no loops (Corollary~\ref{cor:no-loops-funnel})
 
    When a node lies on a cycle, the strict funnel inequality holds, and so the funnel theorems leads to 
    inequalities. 
    Nevertheless, the funnel theorems can be used to solve the master equations explicitly
 on infinite networks that have an infinite number of loops.  Here, 
 the idea is to use the upper bound of the funnel inequality to show that
 the effect of cycles goes to zero as the number of nodes becomes infinite. 
 Therefore, one can effectively use the funnel inequality, even though the network contains numerous loops. 
 This approach has been recently used to compute explicitly the expected adoption/infection   level on infinite
regular networks and on infinite sparse Erd\H{o}s-R\'enyi networks~\cite{ER-25}.

\subsection{Circular networks}
\label{subsec:homogeneous_circles}

In this section, we use the funnel theorem to derive a novel 
inequality  for the expected adoption/infection on circular networks,
 which is of interest by itself, and
will also be used in the proof of Theorem~\ref{thm:f_line^one-sided<f_line_two-sided}.
%
Let $f^{\overrightarrow{\rm circle}}(t)$ denote the expected adoption/infection level in a homogeneous one-sided circle with $M$~nodes, where the node number increases in the counter-clockwise direction, and  
each individual is only influenced by their left neighbor, i.e., 
\begin{equation}
	\label{eq:p_j_q_j_one-sided-circle}
	I_j^0\equiv I^0, \quad 
	p_j\equiv p,\quad 
	q_{k,j}= 
	q \, \mathbbm{1}_{(j-k)\, {\rm mod} \,  M=1},
	\qquad j,k \in {\cal M}.
\end{equation}
Similarly, denote by $f^{\rm circle}(t)$ the expected adoption/infection level  in a homogeneous two-sided circle with
$M$~nodes, where each node can be influenced by its left and right neighbors, i.e.,
\begin{equation}
	\label{eq:p_j_q_j_two-sided-circle}
	I_j^0\equiv I^0, \quad 
p_j\equiv p,\qquad 
q_{k,j}= 
q^{\rm L} \, \mathbbm{1}_{(j-k)\, {\rm mod} \,  M=1}+
q^{\rm R} \, \mathbbm{1}_{(j-k)\, {\rm mod} \,  M=-1},
\qquad j,k \in {\cal M}.
\end{equation}	
	When $q=q^{\rm L}+q^{\rm R}$, 
	the expected adoption/infection levels on one-sided and two-sided  circles are identical~\cite{OR-10}, i.e.,
	\begin{equation}
		\label{eqs:compare_Fibich_Gibori}
		f^{\overrightarrow{\rm circle}}(t; p,q=q^{\rm L}+q^{\rm R},I^0,M)\equiv f^{\rm circle}(t;p,q^{\rm R},q^{\rm L},I^0,M), \qquad t \ge 0.
	\end{equation}
%
%
%
We can use the funnel inequality to derive the following inequality:
	
		\begin{lemma}
		\label{lem:alpha_q-circle-II}
		Let $
		[S^{\overrightarrow{\rm circle}}]:= 1-f^{\overrightarrow{\rm circle}}
		$ denote the expected
		nonadoption/noninfection  level in the Bass/{\rm SI} model~\eqref{eqs:Bass-SI-models-ME} on the one-sided circle~\eqref{eq:p_j_q_j_one-sided-circle}. 
		Let~$q_1,q_2>0$ and 
		   $3 \le M< \infty$. Then for $ t>0$,
		\begin{equation}
			\label{eq:circle_convex}
			[S^{\overrightarrow{\rm circle}}](t;p,q_1,I^0,M) \, [S^{\overrightarrow{\rm circle}}](t;p,q_2,I^0,M)
			<(1-I^0) e^{-pt} \, [S^{\overrightarrow{\rm circle}}](t;p,q_1+q_2,I^0,M).
		\end{equation}
	\end{lemma}
	\begin{proof}
		Consider the Bass/SI model~\eqref{eqs:Bass-SI-models-ME} on the two-sided anisotropic circle~\eqref{eq:p_j_q_j_two-sided-circle}, 
		where the weights of the clockwise and counter-clockwise edges are~$q^{\rm R}:= q_1$ and~$q^{\rm L} := q_2$, respectively.\,\footnote{text}  Let $j \in \{2, \dots, M-1\}$, 
		$A_1 := \left\{1,\ldots, j-1\right\}$, and~$A_2: = \left\{j+1,\ldots,M\right\}$.
		Then $\{A_1,A_2,\{j\}\}$ is a partition of the nodes. The corresponding networks~${\cal N}^{A_1, p_j}$ and~${\cal N}^{A_2, p_j}$ are obtained from the original circle by deleting the edges~$j  \leftarrow j+1$ and~$j-1 \rightarrow j$, respectively
		(see Definition~\ref{def:networks-N^A,N^B}). Since the circle is two-sided, any node~$m \in {\cal M} \setminus \{ j \}$ is influential to~$j$ in both~${\cal N}^{A_1, p_j}$ and~${\cal N}^{A_2, p_j}$. Consequently, 
		$j$~is {\em not}  a funnel node of~$A_1$ and~$A_2$. Therefore, the strict funnel  inequality~\eqref{eq:funnel_strict_inequality-A-p} holds.
		
		The original network is a two-sided circle. Hence, by the equivalence of one-sided and two-sided circles, see~\eqref{eqs:compare_Fibich_Gibori}, 	\begin{subequations}
			\label{eqs:circle1-4}
			\begin{equation}
				\label{eq:circle1}
				[S_j]
				= [S^{\overrightarrow{\rm circle}}](t;p,q_1+q_2, I^0, M).
			\end{equation}
			The calculation of~$[S_j^{A_1,p_j}]$ is as follows.
			In network~${\cal N}^{A_1,p_j}$, we removed the edge~$j  \leftarrow  j+1$. As a result, all the clockwise edges $\{k \leftarrow k+1\}_{k \not=j}$ become non-influential to~$j$. Hence, by the indifference principle, 
			we can compute~$[S_j^{A_1,p_j}]$ on the
			counter-clockwise one-sided circle with~$q^{\rm R}=q_1$, i.e.,
			\begin{equation}
				\label{eq:circle3}
				[S_j^{A_1,p_j}] = [S^{\overrightarrow{\rm circle}}](t;p,q_1, I^0, M).
			\end{equation}
			Similarly,
			\begin{equation}
				\label{eq:circle4}
				[S_j^{A_2,p_j}]= [S^{\overrightarrow{\rm circle}}](t;p,q_2, I^0, M).
			\end{equation}
			By~\eqref{eq:funnel_p_j}.
			\begin{equation}
				\label{eq:circle2}
				[S_j^{p_j}] = (1-I^0) e^{-pt}.
			\end{equation}
		\end{subequations}
		Substituting expressions~\eqref{eqs:circle1-4}
		into the strict funnel  inequality~\eqref{eq:funnel_strict_inequality-A-p} completes the proof. 
	\end{proof}

	The proof of Lemma~\ref{lem:alpha_q-circle-II} shows that 
	inequality~\eqref{eq:circle_convex} is strict, because 
	on the finite circle, $j$~is not a funnel node of 
	$A = \left\{1,\ldots, j-1\right\}$ and~$B = \left\{j+1,\ldots,M\right\}$, 
	and so the strict funnel inequality holds.
	If we let $M \to \infty$, however,
	removing node~$j$ from the network makes the two sets
	$A = \left\{1,\ldots, j-1\right\}$ and~$B = \left\{j+1,\ldots,\infty \right\}$ 
	disconnected.
	Therefore,  on the infinite circle, $j$~is a funnel node of~$A$ and~$B$ 
	(Lemma~\ref{lem:j-is-funnel}). Hence, on the infinite circle the funnel equality~\eqref{eq:funnel_equality-A-p} holds, and so inequality~\eqref{eq:circle_convex}
	becomes an equality as~$M \to \infty$:
	\begin{lemma}
		Let  $[S^{\rm 1D}]:=\lim_{M \to \infty} [S^{\overrightarrow{\rm circle}}](t;p,q,I^0,M)$. Then
		\begin{equation}
			\label{eq:S_1D_q1_q2}
			[S^{\rm 1D}](t;p,q_1,I^0) \, [S^{\rm 1D}](t;p,q_2,I^0)
			=(1-I^0) e^{-pt} [S^{\rm 1D}](t;p,q_1+q_2,I^0), \qquad t \ge 0.
		\end{equation}
	\end{lemma}

	\subsection{Bounded lines}
	\label{sec:bounded-lines}

	Consider the Bass/SI model~\eqref{eqs:Bass-SI-models-ME} on the one-sided line~$\overrightarrow{[1, \dots, M]}$, where each node can only be influenced by its left neighbor, i.e., 
	\begin{equation}
		\label{eq:p_j_q_j_onesided_line}
	I_j^0\equiv I^0, \quad 
p_j\equiv p,\quad 
q_{k,j}= 
q \, \mathbbm{1}_{j-k=1},
\qquad j,k \in {\cal M},
	\end{equation}
and on the two-sided anisotropic line $[1, \dots, M]$,
where each node can be influenced by its left and right neighbors
at the rates of~$q^{\rm L}$  and~$q^{\rm R}$, respectively, i.e.,
\begin{equation}
	\label{eq:p_j_q_j_twosided_line}
	I_j^0\equiv I^0, \quad 
	p_j\equiv p,\qquad 
	q_{k,j}= 
	q^{\rm L} \, \mathbbm{1}_{j-k=1}+
	q^{\rm R} \, \mathbbm{1}_{j-k=-1},
	\qquad j,k \in {\cal M}.
	\qquad 	\qquad k,j\in {\cal M}.
\end{equation} 
	Previously, we used the indifference principle (Theorem~\ref{thm:Indifference}) to derive an explicit expression for the adoption/infection probability of node~$j$ on the one-sided line:
		\begin{lemma}[\cite{Bass-boundary-18}]
			\label{lem:fj_one-sided_line}
			Let $f_j^{\overrightarrow{[1, \dots, M]}}$ denote  
			the adoption/infection probability of node~$j$ in the Bass/{\rm SI} model~\eqref{eqs:Bass-SI-models-ME} on the one-sided line~\eqref{eq:p_j_q_j_onesided_line}. 
			Then
			\begin{equation}
				\label{eq:f_j^onesided=f_circle(j)}
				f_j^{\overrightarrow{[1, \dots, M]}}(t;p,q,I^0)= f^{\rm circle}(t;p,q,I^0,j),
			\end{equation}
		where $f^{\rm circle}(\cdot,j)$ is the expected adoption/infection level on a circle with $j$~nodes, see~\eqref{eq:p_j_q_j_one-sided-circle}.
		\end{lemma}
		In~\cite{Bass-boundary-18}, we also obtained an explicit expression for
		the adoption probability of nodes on the two-sided line. 
		That expression, however, was very cumbersome.
		In this section we use the funnel theorem to obtain a much simpler expression, as follows.
		Let us denote the adoption/infection probability of node~$j$ in a two-sided line by~$f_j^{[1, \dots, M]}$, 
		and its nonadoption probability by 
		$[S_j^{[1, \dots, M]}]:=1-f_j^{[1, \dots, M]}$.
		Let~$[S_j^{\rm L}]$ ($[S_j^{\rm R}]$) denote the nonadoption probability of~$j$ when we discard the influences of all the right (left) neighbors
		by setting $q^{\rm R} \equiv0$ ($q^{\rm L} \equiv0$) in~\eqref{eq:p_j_q_j_twosided_line}, so that the network becomes a left-going (right-going) one-sided line. 		
		\begin{lemma} [\cite{DCDS-23}]
			\label{lem:1D2SWithP_R-hom}
			Consider the Bass/{\rm SI} model~\eqref{eqs:Bass-SI-models-ME} on the two-sided line \eqref{eq:p_j_q_j_twosided_line}.
			Then 
			\begin{equation}
				\label{eq:EquationDiscrete1D2Sided_1}
				[S_j^{[1, \dots, M]}](t)
				=
				\frac{[S_j^{\rm L}](t) \, [S_j^{\rm R}](t)}{(1-I^0)e^{-p t}}, \qquad j\in\mathcal{M}, \qquad t \ge 0.
			\end{equation}
		\end{lemma}
		
		\begin{proof}
			This result was first proved in~\cite{DCDS-23}. 
		Here we provide a simpler proof, by making use of the funnel equality. Let $j$ be an interior node.  Let~$A_1 := \left\{k \in {\cal M} \mid k<j \right\}$ and~$A_2 := \left\{k \in {\cal M} \mid k>j \right\}$. Then $\{A_1,A_2,\{j\}\}$ is a partition of the nodes, and~$j$ is  a  {\em vertex  cut}
		between~$A_1$ and~$A_2$ (see Definition~\ref{def:vertex-cut}). 
		Hence, $j$~is a funnel node of~$A_1$ and~$A_2$ (Lemma~\ref{lem:j-is-funnel}). 
		Therefore, by the funnel  equality~\eqref{eq:funnel_equality-A-p}, 
		\begin{equation}
			\label{eq:funnel_thrm2-2-sided}
			\big [S_j^{{}{[1, \dots, M]}} \big]=
			\frac{[S^{A_1,p_j}_j] \, [S^{A_2,p_j}_j]}
			{[S^{p_j}_j]}	  	.
		\end{equation} 
		In the network~${\cal N}^{A_1,p_j}$, see Definition~\ref{def:more-networks}, the edge $j  \leftarrow  j+1$ is deleted. Therefore, 
		by the indifference principle, 
		$
		[S^{A_1,p_j}_j] =[S_j^{\rm L}]. 
		$	
		Similarly, 
		$
		[S^{A_2,p_j}_j]= [S_j^{\rm R}]. 
		$	
		As always, 
		$
		[S^{p_j}_j] = (1-I^0)e^{-pt}.
		$
		Hence, \eqref{eq:EquationDiscrete1D2Sided_1}~follows.
		
		Relation~\eqref{eq:EquationDiscrete1D2Sided_1} also holds at the boundary nodes.
		Indeed,  the left boundary node  $j=1$ cannot be influenced by any node from the left, and so $[S_{j=1}^{\rm L}] = (1-I^0) e^{-pt}$.
		In addition, by the indifference principle, 
		$[S_{j=1}^{\rm R}]=[S_{j=1}^{{}{[1, \dots, M]}}]$. 
		Therefore, \eqref{eq:EquationDiscrete1D2Sided_1} holds for $j=1$. 
		The proof for the right boundary point is similar. 
		\end{proof}
		
		An explicit expression for the adoption probability~$f_j^{[1, \dots, M]}$ 
		of nodes on a two-sided line with $M$~nodes
		was previously obtained in~\cite{Bass-boundary-18} for the Bass model on the isotropic line. Thus,  $f_j^{[1, \dots, M]} = 1-[S_j^{[1, \dots, M]}]$,
		where 
		\begin{flalign*}
			\label{eq:S_two_sided-old}
			[S_j^{[1, \dots, M]}](\cdot,q^{\rm R}=\frac{q}{2},q^{\rm L}=\frac{q}{2}) = 
			\begin{cases}
				[S^{\overrightarrow{\rm circle}}](\cdot,\frac{q}{2},M),\quad &\text{$j=1, M$,} \\
				e^{-(p+q)t}\left(1+\frac{q}{2}V_j(t)\right), \quad &\text{$j=2, \dots, M-1$,}
			\end{cases}
		\end{flalign*}
		and
		$		V_j\left(t\right) = \int_{0}^{t}e^{(p+q)\tau}\Big[[S^{\overrightarrow{\rm circle}}](\cdot\frac{q}{2},j)[S^{\overrightarrow{\rm circle}}](\cdot\frac{q}{2},M-j)+ [S^{\overrightarrow{\rm circle}}](\cdot\frac{q}{2},j-1)[S^{\overrightarrow{\rm circle}}](\cdot\frac{q}{2},M-j+1)\Big]d\tau.
		$
		A simpler expression, which is also valid in the anisotropic case
		~$q^{\rm L}\not=q^{\rm R}$,  can be obtained using the funnel equality:
		\begin{theorem}
			\label{lem:fj_twosided_new}
			Consider the Bass/{\rm SI} model~\eqref{eqs:Bass-SI-models-ME} on the two-sided line \eqref{eq:p_j_q_j_twosided_line}.	Then 
			\begin{equation}
			\label{eq:S_two_sided}
			[S_j^{{}{[1, \dots, M]}}](t;p,q^{\rm R},q^{\rm L},I^0) =
			\frac{[S^{\overrightarrow{\rm circle}}](t;p,q^{\rm L},I^0,j) \, 
				[S^{\overrightarrow{\rm circle}}](t;p,q^{\rm R},I^0,j^*)}
			{(1-I^0)e^{-pt}} , \qquad j \in {\cal M}, 
		\end{equation}
		where  
		\begin{equation}
			\label{eq:j*-2-sided-line}
			j^* := M+1-j
		\end{equation} 
	is the number of nodes from $M$ to $j$ (and thus the symmetric node  to~$j$ about the midpoint).
			\end{theorem}
		\begin{proof}
		By Lemma~\ref{lem:1D2SWithP_R-hom}, 
		$
		[S_j^{[1, \dots, M]}]
		=
		\frac{[S_j^{\rm L}] \, [S_j^{\rm R}]}{(1-I^0)e^{-p t}}.
		$
		On the two-sided line $[1, \dots, M]$,
		$$
		[S_j^{\rm L}](t) = [S_j^{\overrightarrow{[1, \dots, M]}}](t;p,q^{\rm L},I^0), \qquad 
		[S_j^{\rm R}](t) = [S_{j}^{\overleftarrow{[1, \dots, M]}}](t;p,q^{\rm R},I^0) = [S_{M-j+1}^{\overrightarrow{[1, \dots, M]}}](t;p,q^{\rm R},I^0).
		$$
		Since
		$
		[S_j^{\overrightarrow{[1, \dots, M]}}](\cdot) = [S^{\overrightarrow{\rm circle}}](\cdot,j),
		$
		see~\eqref{eq:f_j^onesided=f_circle(j)}, the result follows. 
		\end{proof}

		\subsubsection{$f^{\protect\overrightarrow{[1, \dots, M]}}<f^{{[1, \dots, M]}}$} 
		\label{sec:f_one-sided<f_two-sided}
		
		As noted, when~$q = q^{\rm R}+q^{\rm L}$, the expected adoption/infection levels on the 
		one-sided and two-sided circles are identical, i.e., 
		$f^{\overrightarrow{\rm circle}} \equiv f^{\rm circle}$, see \eqref{eqs:compare_Fibich_Gibori}.
		On finite lines, however, this is not the case. Indeed, in~\cite{Bass-boundary-18} we showed that one-sided spreading is strictly
		slower that isotropic two-sided spreading 
		i.e., 
		$$
		f_{\rm line}^{\overrightarrow{[1, \dots, M]}}(\cdot,q)<
		f^{{[1, \dots, M]}}\left(\cdot,q^{\rm L}= \frac{q}{2},q^{\rm R}= \frac{q}{2}\right), \qquad t>0.
		$$
		The availability of the new explicit expression~\eqref{eq:S_two_sided}
		for~$[S_j^{[1, \dots, M]}]$ 
		allows us to generalize this result to the anisotropic two-sided case
		($q^{\rm R} \neq q^{\rm L}$) and also to provide a much simpler and shorter proof: 
		\begin{theorem}
			\label{thm:f_line^one-sided<f_line_two-sided}
			Let  $q = q^{\rm L} + q^{\rm R}$. Then for any $q_L,q_R>0$ and $2 \le M <\infty$, 
		$$
	f_{1}^{\overrightarrow{[1, \dots, M]}}(t;p,q,I^0)<f_{1}^{{}{[1, \dots, M]}}(t;p,q^{\rm L},q^{\rm R},I^0), \qquad t>0  .
	$$
		\end{theorem}
	\begin{proof}
	Let $t>0$.
Since $f = \frac1M \sum_{j=1}^{M}f_{j}$,  
we need to prove that
$	\sum_{j=1}^{M}f^{\overrightarrow{[1, \dots, M]}}_j< \sum_{j=1}^{M} f^{{}{[1, \dots, M]}}_j$.   
The key to proving this inequality is to show that it holds for {\em  any pair of nodes $\{j,j^*\}$ that are symmetric about the midpoint}, 
see~\eqref{eq:j*-2-sided-line}
i.e., that
\begin{equation*}
	f^{\overrightarrow{[1, \dots, M]}}_j
	+f^{\overrightarrow{[1, \dots, M]}}_{j^*}
	<
	f^{{}{[1, \dots, M]}}_j
	+f^{{}{[1, \dots, M]}}_{j^*}, \qquad j \in  {\cal M} .
\end{equation*}
%
This inequality can be rewritten as   
$$
[S^{\overrightarrow{[1, \dots, M]}}_j]+[S^{\overrightarrow{[1, \dots, M]}}_{j^*}]> [S^{{}{[1, \dots, M]}}_j]+[S^{{}{[1, \dots, M]}}_{j^*}], \qquad j \in  {\cal M} .
$$
By~\eqref{eq:f_j^onesided=f_circle(j)} and~\eqref{eq:S_two_sided}, it suffices to prove for $t>0$ that
\begin{equation}
	\label{eq:apolo}
	\begin{split}
		\lefteqn{[S^{\overrightarrow{\rm circle}}](t;p,q,I^0,j)+ 
			[S^{\overrightarrow{\rm circle}}](t;p,q,I^0,j^*)
			>}  		
		\\[5px] 
		& \quad
		\frac{[S^{\overrightarrow{\rm circle}}](t;p,q^{\rm L},I^0,j) \,  [S^{\overrightarrow{\rm circle}}](t;p, q^{\rm R},I^0,j^*)}{(1-I^0)e^{-pt}} 
		+	\frac{[S^{\overrightarrow{\rm circle}}](t;p, q^{\rm L},I^0,j^*) \, [S^{\overrightarrow{\rm circle}}](t;p, q^{\rm R},I^0,j)}{(1-I^0)e^{-pt}}.
	\end{split}
\end{equation}

Let $s(q,j):=[S^{\overrightarrow{\rm circle}}](t; p, q,I^0,j)$.
Then~\eqref{eq:apolo} reads
$$
s(q,j)+ 
s(q,j^*)
>	
\frac{[s(q^{\rm L},j) \, s(q^{\rm R},j^*)}{(1-I^0)e^{-pt}} 
+	\frac{s(q^{\rm L},j^*) \, s(q^{\rm R},j)}{(1-I^0)e^{-pt}}.
$$
By Lemma~\ref{lem:alpha_q-circle-II}, 
$$
s(q,j) >  
\frac{s(q^{\rm R},j)\, s(q^{\rm L},j)}{(1-I^0)e^{-pt}},  
\qquad s(q,j^*) > 
\frac{s(q^{\rm R},j^*)\, s(q^{\rm L},j^*)}{(1-I^0)e^{-pt}}.
$$
Therefore, to prove~\eqref{eq:apolo}, it suffices to show that
\begin{align*}
	&s(q^{\rm L},j) \, s(q^{\rm R},j)
	+
	s(q^{\rm L},j^*)\, s(q^{\rm R},j^*) 
	>
	s(q^{\rm L},j) \, s(q^{\rm R},j^*)
	+
	s(q^{\rm L},j^*) \, s(q^{\rm R},j), 
\end{align*}
i.e., that 
\begin{equation}
\label{eq:(s-s)(s-s)-ineqaulity}
\Big(s(q^{\rm L},j)-s(q^{\rm L},j^*)\Big)
\Big(s(q^{\rm R},j)- s(q^{\rm R},j^*)\Big)>0.
\end{equation}
This inequality follows from the strict monotonicity of 
$s(q,j):=[S^{\overrightarrow{\rm circle}}](t; p, q,I^0,j)$
in~$j$, see~\cite{Bass-monotone-convergence-23}.\,\footnote{
	In fact, ~\eqref{eq:(s-s)(s-s)-ineqaulity} is a strict inequality 
	for $j\not=j^*$ and an equality for $j=j^* = \frac{M+1}2$. That is not a problem, however, since only the sum over all pairs needs to satisfy a strict inequality.} Therefore, we proved~\eqref{eq:apolo}. 
\end{proof}

		\subsection{Multi-dimensional Cartesian networks}
		\label{sec:f_1D-lower-bound}
		
		Consider an infinite $n$-dimensional Cartesian network~$\mathbb{Z}^{n}$, where nodes are labeled by their $n$-dimensional coordinate vector~${\bf j} = (j_1,\cdots,j_n) \in \mathbb{Z}^{n}$.
			Each node can be influenced by its $2n$ nearest-neighbors at the rate of~$\frac{q}{2n}$, and so the external and internal parameters are
			\begin{equation}
				\label{eq:Bass-model-D-dimensional}
					I_{\bf j}^0 \equiv I^0, \quad 
					p_{\bf j} \equiv p, \quad q
				_{{\bf k},{\bf j}} = \left\{ 
				\begin{array}{ll}
					\frac{q}{2n}, & \quad {\bf j} - {\bf k} = \pm {\bf \widehat{e}}_i ,\\
					0, & \quad  \mbox{otherwise},
				\end{array}
				\right. ,\qquad {\bf k}, {\bf j} \in \mathbb{Z}^{n}, \quad i \in \left\{1,\ldots, n\right\}. 
			\end{equation}
		We denote the expected adoption/infection  on~$\mathbb{Z}^{n}$ 
		by~$f^{n \rm  D}$.
		In~\cite{OR-10}, it was observed numerically that 
		$f^{n \rm  D}$ is monotonically increasing in~$n$, i.e., 
		$$
		f^{\rm 1D}(t;p,q,I^0)<f^{\rm 2D}(t;p,q,I^0) <f^{\rm 3D}(t;p,q,I^0) < \cdots,   \qquad  t>0. 
		$$ 
	So far, however, 
		this result was only proved for small times~\cite[Lemma~14]{MOR-23}.
		We can use the funnel theorem to provide a partial proof, 
		namely,  that
		$f^{n \rm  D}>f^{\rm 1D}$ for all~$n\geq 2$:
		\begin{lemma} 
			\label{lem:f_1D_bound}
			%
			For any~$n\geq 2$ and $t>0$, $f^{n \rm  D}(t;p,q,I^0) > f^{\rm 1D}(t;p,q,I^0)$.
		\end{lemma}
		\begin{proof}
		Let $n\geq 2$.
		Denote  the origin node by ${\bf 0}:=(0, \dots, 0) \in \mathbb{Z}^{n}$. 
		By  translation invariance, the adoption/infection probability of 
		node~${\bf 0}$ is
		\begin{equation}
			\label{eq:N1_cart}
			f^{\mathbb{Z}^{n}}_{\bf 0}(t)  = f^{n \rm  D}(t).
		\end{equation}
		Let network~$\mathbb{Z}^{n}_{\rm rays}$ be obtained from~$\mathbb{Z}^{n}$ by removing all the edges, except for those that lie on lines that go through~${\bf 0}$ and also point towards~${\bf 0}$. 
		Hence, the node~${\bf 0}$ in the network~$\mathbb{Z}^{n}_{\rm rays}$ is the intersection point  
		of~$2n$ one-sided semi-infinite rays with edge weights~$\widetilde{q} = \frac{q}{2n}$. 
				In Lemma~\ref{lem:intersection} below we will prove that  
		\begin{equation}
			\label{eq:N1_1D}
			f^{\mathbb{Z}^{\rm n}_{\rm rays}}_{\bf 0}(t)
			=f^{\rm 1D}(t). 
		\end{equation}
		Since some of the edges that were removed in~$\mathbb{Z}^{n}_{\rm rays}$ are influential to node~${\bf 0}$, 
		by the strong dominance principle for nodes~\cite{Bass-boundary-18}, we have that
		\begin{equation}
			\label{eq:N1>N2}
			f^{\mathbb{Z}^{\rm n}}_{\bf 0}(t) 
			>
			f^{\mathbb{Z}^{\rm n}_{\rm rays}}_{\bf 0}(t).
		\end{equation}
		Combining~\eqref{eq:N1_cart}, \eqref{eq:N1_1D}, and \eqref{eq:N1>N2} gives the result.
	\end{proof}
		
\begin{figure}[ht!]
			\begin{center}
				\scalebox{0.8}{\includegraphics{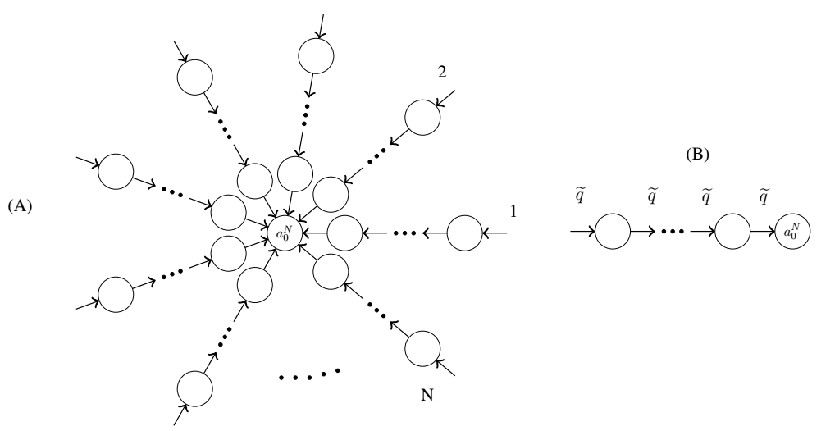}}
				\caption{(A)~Node $a_0^N$ is at the intersection of~$N$ one-sided semi-infinite rays. (B)~A single semi-infinite ray. The weight of all edges is~$\widetilde{q}$}
				\label{fig:intersection}
			\end{center}
		\end{figure}

		To finish the proof of Lemma~\ref{lem:f_1D_bound}, we prove
		\begin{lemma}
			\label{lem:intersection}
			Let node~$a_0^{N}$ be the intersection of~$n$ identical one-sided semi-infinite rays, such that the weight of all edges is~$\widetilde{q}$ {\rm (}see Figure~{\rm \ref{fig:intersection})}. Then
				\begin{equation}
					\label{eq:intersection}
			\mathbb{P}(X_{a_0^{N}}(t)=1) = f^{\rm 1D}(t;p,N\widetilde{q}).
				\end{equation}
		\end{lemma}

	\begin{proof}
		We proceed by induction on~$N$, the number of rays. 
	%
			By Lemma~\ref{lem:fj_one-sided_line},
		$f_{j}^{\overrightarrow{[j-M+1, \dots, j]}}(t) =	f^{\rm circle}(t;M)$.
		Therefore, 
		\begin{equation}
			\label{eq:basecase}
			\begin{aligned}
				f_j^{\overrightarrow{(-\infty, \dots, j]}}(t)
				=
				\lim_{M \to \infty} f_{j}^{\overrightarrow{[j-M+1, \dots, j]}}(t)
				= 	\lim_{M \to \infty}
				f^{\rm circle}(t;M) 
				= f^{\rm 1D}(t),
			\end{aligned}
		\end{equation}
		where the last equality follows from~\cite{OR-10}.
		Therefore, we proved the induction base $N=1$.

		For the induction stage, we prove the equivalent result 
		\begin{equation}
			\label{eq:intersection-S}
			[S_{a_0^{N}}]  = [S^{\rm 1D}](t;p,N\widetilde{q}),
		\end{equation}
		where $[S_{a_0^{N}}]:= 1-f_{a_0^{N}}$ and $[S^{\rm 1D}] = 1-f^{\rm 1D}$.
		Thus, we assume that~\eqref{eq:intersection-S} holds for
		$N$~rays, and prove that it also holds for~$N+1$ rays. 
		Indeed, let~$A_1$ denote the first~$N$ rays, and let~$A_2$ denote the $(N+1){\rm th}$~ray. Since the node~$a_0^{N+1}$ is a {\em vertex cut} of~$A_1$ and~$A_2$ (Definition~\ref{def:vertex-cut}), 
		it is a funnel node of~$A_1$ and~$A_2$ (Lemma~\ref{lem:j-is-funnel}). Therefore, by the funnel equality~\eqref{eq:funnel_equality-A-p},
		\begin{equation}
			\label{eq:induction0}
			\left[S_{a_0^{N+1}}\right]  = \frac{	\left[S^{A_1,p}_{a_0^{N+1}}\right] 	\left[S^{A_2,p}_{a_0^{N+1}}\right]}{	\left[S^{p}_{a_0^{N+1}}\right]}.
		\end{equation}
		By construction, $	\left[S^{A_1,p}_{a_0^{N+1}}\right] = \left[S_{a_0^{N}}\right]$. 
		Hence, by the induction assumption, see~\eqref{eq:intersection-S},
		\begin{subequations}
			\label{eqs:induction123}
			\begin{equation}
				\label{eq:induction1}
				\left[S^{A_1,p}_{a_0^{N+1}}\right] = [S^{\rm 1D}](t;p,N\widetilde{q}).
			\end{equation} 
			Similarly, by construction, $	\left[S^{A_2,p}_{a_0^{N+1}}\right] = \left[S_{a_0^{N=1}}\right]$. 
			Therefore, by~\eqref{eq:basecase}, 	\begin{equation}
				\label{eq:induction2}
				\left[S^{A_2,p}_{a_0^{N+1}}\right]
				= [S^{\rm 1D}](t;p,\widetilde{q}).
			\end{equation}
			By~\eqref{eq:funnel_p_j}, 
			\begin{equation}
				\label{eq:induction3}
				\left[S^{p}_{a_0^{N+1}}\right] = (1-I^0)e^{-pt}.
			\end{equation}
		\end{subequations}
		Substituting the expressions~\eqref{eqs:induction123} 
		in~\eqref{eq:induction0} and using~\eqref{eq:S_1D_q1_q2} 
		gives
		\begin{equation*}
			\begin{aligned}
				\left[S_{a_0^{N+1}}\right]  = [S^{\rm 1D}](t;p,(N+1)\widetilde{q}),
			\end{aligned}
		\end{equation*}
		as desired.
	\end{proof}

		 \section{Proof of Theorems~\ref{thm:[Si,Sj]>[Si][Sj]-K} and~\ref{thm:[S_Omega]-prod[S_i]-N-paths}}   
		  \label{sec:Sij>SiSj-proof}
		  
		\subsection{Preliminary results}
		
		We first note some consequences of the master equations:
		
		\begin{lemma}
			\label{lem:[S_{Omega}^0]>0}
			$[S_{\Omega}^0]>0$ for all $\emptyset \not=\Omega \subset \cal M$. 
		\end{lemma}
		\begin{proof}
			This follows from~\eqref{eq:parameters-assumtions-A}   and~\eqref{eq:master-eqs-genera-icl}.  
		\end{proof}
		
		\begin{lemma}
			\label{lem:[S_Omega]<=[S_Omega^0]e^-pt}
			Let $\emptyset \not=\Omega \subset \cal M$. 
			Then 
			\begin{equation}
				\label{eq:[S]<e^-pt}
				0<[S_{\Omega}^0] e^{-\big( p_\Omega	+\sum_{k \in \Omega^{\rm c}} q_{k,\Omega}\big)t} \le 
				[S_{\Omega}] \le [S_{\Omega}^0] e^{-p_\Omega t}<1, \qquad t>0.
			\end{equation}
		\end{lemma}
		\begin{proof}
			The master equation~\eqref{eq:master-eqs-general} can be rewritten as
					\begin{equation}
				\label{eq:master-eqs-general-B}
				\frac{d [S_{\Omega}]}{dt}  = 
				- p_\Omega \, [S_{\Omega}] 
				-\sum_{k \in \Omega^{\rm c}} q_{k,\Omega}   \left([S_{\Omega}]- [S_{\Omega,k}]\right).
			\end{equation}
			If the event  $S_{\Omega,k}$ occurs, the event  $S_{\Omega}$ occurs as well.
			Therefore, 
			\begin{equation}
				   \label{eq:[S_Omega]ge[S_Omega,k]}
				[S_{\Omega}]  \ge  [S_{\Omega,k}].
			\end{equation} 
			In addition, we have that $q_{k,\Omega} \ge 0$ and  $[S_{\Omega,k}] \ge 0$. 
			Therefore, from equation~\eqref{eq:master-eqs-general-B}  we have that
			\begin{equation*}
				- p_\Omega \, [S_{\Omega}]
				\ge 				
				\frac{d [S_{\Omega}]}{dt}  \ge  
				-\bigg( p_\Omega
				+\sum_{k \in \Omega^{\rm c}} q_{k,\Omega}\bigg)[S_{\Omega}].
			\end{equation*}
			In addition, $[S_{\Omega}^0]>0$, see Lemma~\ref{lem:[S_{Omega}^0]>0}.
			Therefore, the result follows.
		\end{proof}


		\begin{lemma}\label{lem:influ}
			Let $\emptyset \not=\Omega \subsetneq \cal M$ and $k \in \Omega^c$. Then 
			$[S_{\Omega, k}]< [S_\Omega]$ for $t>0$.
		\end{lemma}
		\begin{proof}
			By the law of sum of probability
			$$
			[S_{\Omega}] -[S_{\Omega, k}] =  [S_{\Omega} \cap I_k]
			 \ge [S_{{\cal M}_{-k}} \cap I_k]
			  = [S_{{\cal M}_{-k}} ]-[S_{\cal M}],
			$$
			where $[S_{\Omega} \cap I_k]:=\mathbb{P}(X_\Omega=0, X_k=1)$
			and ${\cal M}_{-k}:={\cal M} \setminus \{k\}$. 
			Therefore, it is sufficient to prove that $y(t):=[S_{{\cal M}_{-k}} ]-[S_{\cal M}]>0$ for $t>0$.
			From the master equations~\eqref{eqs:master-eqs-general},
			we have that 
					$$
			[S_{\cal M}] = e^{-p_{\cal M} t}\prod_{m \in \cal M} (1-I_m^0), 
			$$ 
			and 			
			\begin{equation}
			\label{eq:dydt-[S_Omega]-[S_Omega,k]}
			  \frac{dy}{dt} = -(p_{{\cal M}_{-k}}+q_{k,{\cal M}_{-k}})y + 
			  p_k [S_{\cal M}], \qquad y(0) = I_k^0 \prod_{m \in {{\cal M}_{-k}}}(1-I_m^0).
				\end{equation}
			Therefore,  by~\eqref{eq:parameters-assumtions-A}, $[S_{\cal M}]>0$ for $t>0$ and $y(0) \ge 0$.
			Furthermore, by~\eqref{eq:parameters-assumtions-B}, either $p_k>0$ or $y(0)>0$. 
			Therefore, it follows from~\eqref{eq:dydt-[S_Omega]-[S_Omega,k]} that $y(t)>0$   for $t>0$.
			\end{proof}

		\subsection{Proof of Theorem~\ref{thm:[Si,Sj]>[Si][Sj]-K}}
		\label{sec:[Si,Sj]>[Si][Sj]-K}
		
		Before proving Theorem~\ref{thm:[Si,Sj]>[Si][Sj]-K}, some auxiliary results will be needed. 
		\begin{lemma}
			Let $\emptyset \not= \Omega_1,\Omega_2 \subset \cal M$ such that 
			$\Omega_1 \cap \Omega_2 = \emptyset$. Denote 
			\begin{equation}
				\label{eq:Qdef}
				Q_{\Omega_1,\Omega_2}\eqdef [S_{\Omega_1,\Omega_2}]-[S_{\Omega_1}][S_{\Omega_2}].
			\end{equation}
			Then $	Q_{\Omega_1,\Omega_2}(t
			)$ satisfies the equation
			\begin{subequations}
				\label{eqs:dtQ}
				\begin{equation}
					\label{eq:dtQ}
					\begin{aligned}
						\frac{d Q_{\Omega_1,\Omega_2}}{dt} & + \Big( p_{\Omega_1 \cup \Omega_2}+
						\sum_{m\notin \Omega_1\cup \Omega_2} q_{m, \Omega_1\cup \Omega_2} \Big) Q_{\Omega_1,\Omega_2}
						\\&\quad= \sum_{m\notin \Omega_1\cup \Omega_2}\Big(q_{m,\Omega_1} Q_{\Omega_1\cup\{m\},\Omega_2}+
						q_{m,\Omega_2} Q_{\Omega_1,\Omega_2\cup\{m\}}\Big)
						\\&\qquad + \sum_{m\in \Omega_2}  q_{m,\Omega_1} \Big( [S_{\Omega_1}]-[S_{\Omega_1,m}]\Big) [S_{\Omega_2}] 
						+\sum_{m\in \Omega_1}  q_{m,\Omega_2} \Big( [S_{\Omega_2}]-[S_{\Omega_2,m}]\Big) [S_{\Omega_1}], 	
					\end{aligned}
				\end{equation}
				subject to the initial condition
				\begin{equation}
					Q_{\Omega_1,\Omega_2}(0)=0. 
				\end{equation}
			\end{subequations}
		\end{lemma}
		\begin{proof}
			Using~\eqref{eq:Qdef} and  the master equations~\eqref{eq:master-eqs-general}, we have that  
			\begin{equation*}
				\begin{aligned}
					\frac{d Q_{\Omega_1,\Omega_2}}{dt} &=\frac{d [S_{\Omega_1,\Omega_2}]}{dt} - [S_{\Omega_1}]\frac{d[S_{\Omega_2}]}{dt} -[S_{\Omega_2}]\frac{d [S_{\Omega_1}]}{dt} 
					\\&=
					-\Big( p_{\Omega_1 \cup \Omega_2}+\sum_{m\notin \Omega_1\cup \Omega_2} q_{m,\Omega_1\cup \Omega_2} \Big) [S_{\Omega_1,\Omega_2}]
					+ \sum_{m\notin \Omega_1\cup \Omega_2} q_{m,\Omega_1\cup \Omega_2} [S_{\Omega_1\cup \Omega_2,m}]
					\\&\quad+[S_{\Omega_1}]\Big(p_{\Omega_2}+\sum_{m\notin  \Omega_2}q_{m,\Omega_2} \Big) [S_{\Omega_2}]
					-[S_{\Omega_1}] \sum_{m\notin  \Omega_2}q_{m,\Omega_2} [S_{\Omega_2,m}]
					\\&\quad+[S_{\Omega_2}]\big(p_{\Omega_1}+\sum_{m\notin  \Omega_1}q_{m,\Omega_1} \Big) [S_{\Omega_1}]
					-[S_{\Omega_2}] \sum_{m\notin  \Omega_1}q_{m,\Omega_1} [S_{\Omega_1,m}].
					\\&=
	-\Big( p_{\Omega_1 \cup \Omega_2}+\sum_{m\notin \Omega_1\cup \Omega_2} q_{m,\Omega_1\cup \Omega_2} \Big)    \big(Q_{\Omega_1,\Omega_2}-[S_{\Omega_1]\,[\Omega_2}] \big)
	\\ & \quad + \sum_{m\notin \Omega_1\cup \Omega_2} q_{m,\Omega_1} \Big(Q_{\Omega_1\cup\{m\},\Omega_2}-
	[S_{\Omega_1,m}] \, [S_{\Omega_2}]\Big)
	\\ & \quad + \sum_{m\notin \Omega_1\cup \Omega_2} q_{m, \Omega_2}\Big(Q_{\Omega_1,\Omega_2\cup\{m\}}-
	[S_{\Omega_1}] \, [S_{\Omega_2,m}]\Big)
	\\&\quad+[S_{\Omega_1}]\Big(p_{\Omega_2}+\sum_{m\notin  \Omega_2}q_{m,\Omega_2} \Big) [S_{\Omega_2}]
	-[S_{\Omega_1}] \sum_{m\notin  \Omega_2}q_{m,\Omega_2} [S_{\Omega_2,m}]
	\\&\quad+[S_{\Omega_2}]\big(p_{\Omega_1}+\sum_{m\notin  \Omega_1}q_{m,\Omega_1} \Big) [S_{\Omega_1}]
	-[S_{\Omega_2}] \sum_{m\notin  \Omega_1}q_{m,\Omega_1} [S_{\Omega_1,m}],
				\end{aligned}
			\end{equation*}
			which leads to~\eqref{eq:dtQ}.
			The initial condition follows from the independence of the initial conditions
			of nodes, see~\eqref{eq:p:initial_cond_uncor-two_sided_line}.
			
		\end{proof}


		\begin{lemma}
			\label{lem:Qnonneg}
			Let $\emptyset \not= \Omega_1,\Omega_2 \subset \cal M$ such that 
			$\Omega_1 \cap \Omega_2 = \emptyset$. Then 
			$Q_{\Omega_1,\Omega_2}(t) \ge 0$ for $t \ge 0$.
		\end{lemma}
		
		\begin{proof}
			We proceed by backwards induction on the size of~$\Omega_1\cup \Omega_2$.
			Consider the induction base where $\Omega_1 \cup \Omega_2 = \cal M$. Then equation~\eqref{eqs:dtQ} for~$Q_{\Omega_1,\Omega_2}$ reduces to
			\begin{subequations}
				\label{eqs:odeall}
				\begin{equation}
					\label{eq:odeall}
					\frac{d Q_{\Omega_1,\Omega_2}}{dt} 
					+c_{\Omega_1,\Omega_2}\, Q_{\Omega_1,\Omega_2}
					=\sum_{k\in \Omega_2}
					q_{k,\Omega_1} \Big( [S_{\Omega_1}]-[S_{\Omega_1,k}]\Big) [S_{\Omega_2}] 
					+\sum_{j\in \Omega_1} q_{j,\Omega_2}\Big( [S_{\Omega_2}]-[S_{\Omega_2,j}]\Big) [S_{\Omega_1}], 
				\end{equation}
				where $c_{\Omega_1,\Omega_2}:= p_{\Omega_1 \cup \Omega_2}+
				\sum_{m\notin \Omega_1\cup \Omega_2} q_{m, \Omega_1\cup \Omega_2} \ge 0$, subject to 
				\begin{equation}
					Q_{\Omega_1,\Omega_2}(0)=0. 
				\end{equation}
			\end{subequations}
			By Lemma~\ref{lem:influ}, 
			$[S_{\Omega_1}]-[S_{\Omega_1,k}] > 0$ and 
			$[S_{\Omega_2}]-[S_{\Omega_2,j}] > 0$. In addition, $q_{k,\Omega_1} \ge 0$   and $q_{j,\Omega_2} \ge 0$. 
			Therefore, we have that 
			\begin{equation}\label{eq:dtQge}
				\frac{d Q_{\Omega_1,\Omega_2}}{dt}+ c_{\Omega_1,\Omega_2} \, Q_{\Omega_1,\Omega_2} \ge 0,  \qquad Q_{\Omega_1,\Omega_2}(0)=0.
			\end{equation}
			This differential inequality implies that $Q_{\Omega_1,\Omega_2}(t) \ge 0$  for $ t \ge 0$. 
			
			Assume by induction that $Q_{\Omega_1,\Omega_2}(t) \ge 0$ for $t \ge 0$  
			for all $\Omega_1,\Omega_2$ for which $|\Omega_1\cup \Omega_2| = n+1$. 
			Consider $\Omega_1,\Omega_2$  for which $|\Omega_1\cup \Omega_2| = n$. 
			Then the right-hand side of equation~\eqref{eq:dtQ} is nonnegative.  Therefore,
			the differential inequality~\eqref{eq:dtQge} holds, and so
			$Q_{\Omega_1,\Omega_2} \ge 0$ for $t \ge 0$.
		\end{proof}

		\begin{lemma}
			\label{lem:[Si,Sj]>[Si][Sj]}
			Consider the Bass/{\rm SI}  model~\eqref{eqs:Bass-SI-models-ME}. 
			Let $\emptyset \not=\Omega_1, \Omega_2 \subset {\cal M}$ such that $\Omega_1 \cap \Omega_2 = \emptyset$. Then 
			\begin{equation}
					\label{eq:[S_Omega_ij]>=[S_Omega_i][S_Omega_j]}
				[S_{\Omega_1 \cup \Omega_2}]  \ge  [S_{\Omega_1}] \, [S_{\Omega_2}], \qquad t  \ge  0.
			\end{equation}
			In addition, 
			\begin{enumerate}
				\item If  there exists a node which is influential to both~$\Omega_1$ and~$\Omega_2$,	then 
				\begin{equation}
					\label{eq:[S_Omega_ij]>[S_Omega_i][S_Omega_j]}
					[S_{\Omega_1 \cup \Omega_2}] > [S_{\Omega_1}] \, [S_{\Omega_2}], \qquad t > 0.
				\end{equation}
				\item If, however, there is no node which is influential to both~$\Omega_1$ and~$\Omega_2$, then 
				\begin{equation}
					\label{eq:[S_Omega_ij]=[S_Omega_i][S_Omega_j]}
					[S_{\Omega_1 \cup \Omega_2}] = [S_{\Omega_1}] \, [S_{\Omega_2}], \qquad t \ge 0.
				\end{equation}
			\end{enumerate}
		\end{lemma}
		\begin{proof}
			Inequality~\eqref{eq:[S_Omega_ij]>=[S_Omega_i][S_Omega_j]} is Lemma~\ref{lem:Qnonneg}.
			To prove~\eqref{eq:[S_Omega_ij]>[S_Omega_i][S_Omega_j]} and~\eqref{eq:[S_Omega_ij]=[S_Omega_i][S_Omega_j]},
			we proceed by backwards induction on
			the size of $\Omega_1\cup \Omega_2$.
			
			Consider the induction base where $\Omega_1\cup \Omega_2  =  \cal M$. Then $Q_{\Omega_1,\Omega_2}$  satisfies eq.~\eqref{eqs:odeall}.
			Therefore, since the right-hand side of~\eqref{eqs:odeall} is non-negative, see~\eqref{eq:dtQge},
			then $Q_{\Omega_1,\Omega_2}>0$ is and only if 
			the right-hand side of~\eqref{eqs:odeall} is positive. 
			%
			By Lemma~\ref{lem:influ},  $[S_{\Omega_1}]-[S_{\Omega_1,k}]>0$ and $[S_{\Omega_2}]-[S_{\Omega_2,j}]>0$ for all $j$ and $k$.
			Hence, $Q_{\Omega_1,\Omega_2}(t)$ is positive for all $t>0$  if and only if  there exist $j\in \Omega_1$ and $k\in \Omega_2$ such
			that either $q_{j,k}>0$ or $q_{k,j}>0$ so that the inhomogeneous term in the ODE~\eqref{eq:odeall} is positive, and is identically zero
			otherwise. This proves the theorem for~$\Omega_1\cup \Omega_2 = \cal M$, since in this case the only relevant paths, see Lemma~\ref{lem:paths-influential-node}, are directed edges from~$\Omega_1$ to~$\Omega_2$ or from~$\Omega_2$ to~$\Omega_1$.
			
			Now assume by induction that the lemma holds for all~$\Omega_1, \Omega_2$ for which $|\Omega_1\cup \Omega_2| = n+1$. Consider $\Omega_1,\Omega_2$  for which $|\Omega_1\cup \Omega_2| = n$. Since $[S_{\Omega_1}]-[S_{\Omega_1,k}]$ and $[S_{\Omega_2}]-[S_{\Omega_2,j}]$ are both positive  (Lemma~\ref{lem:influ}), 
			and $	Q_{\Omega_1\cup\{m\},\Omega_2}$ and $Q_{\Omega_1,\Omega_2\cup\{m\}}$
			are nonnegative by Lemma~\ref{lem:Qnonneg}, 
			equation~\eqref{eqs:dtQ} implies that $Q_{\Omega_1,\Omega_2}>0$ for $t>0$ if and only if  at least one of the following 3 conditions holds, and is identically zero otherwise:
			\begin{enumerate}
				\item[C1.] \label{eq:C1-SiSj} For some $j\in \Omega_1$ and $k\in \Omega_2$, either $q_{j,k}>0$ or $q_{k,j}>0$.
				\item[C2.] For some $m\notin \Omega_1\cup \Omega_2$ and $j\in \Omega_1$, $q_{m,j}>0$ and $Q_{\Omega_1\cup\{m\},\Omega_2}>0$.
				\item[C3.] For some $m\notin \Omega_1\cup \Omega_2$ and $k\in \Omega_2$, $q_{m,k}>0$ and $Q_{\Omega_1,\Omega_2\cup\{m\}}>0$. 
			\end{enumerate}
			Therefore, to finish the proof, we need to show that  at least one of the Conditions C1--C3 holds if and only if   there exist a path of the claimed forms
			in Lemma~\ref{lem:paths-influential-node}.

			We first show if any of Conditions C1--C3 holds, there exists a path  of the claimed form: 
			\begin{itemize}
				\item 
				Assume that Condition~C1 holds. Then there exists a single-edge path
				from~$\Omega_1$ to~$\Omega_2$ or from~$\Omega_2$ to~$\Omega_1$.
				\item 
				Assume that Condition~C2 holds.  Then 
				there is an edge from~$m$ to~$j \in \Omega_1$. In addition, since
				$Q_{\Omega_1\cup\{m\},\Omega_2}>0$, then by the induction hypothesis,
				\begin{itemize}
					\item[D1.]  there is a path from $\Omega_1\cup\{m\}$ to $\Omega_2$, or
					\item[D2.]   there is a path from $\Omega_2$ to
					$\Omega_1\cup\{m\}$, or
					\item[D3.] there is a node $\tilde{m} \notin \Omega_1\cup\{m\}\cup \Omega_2$ and paths from~$\tilde{m}$ to~$\Omega_1\cup \{m\}$ and to~$\Omega_2$.
				\end{itemize}
				Now, 
				\begin{itemize} 
					\item If Condition~D1 holds, there is either a path from $\Omega_1$ to $\Omega_2$, or there are paths from $m$ to $\Omega_1$ and to $\Omega_2$. 
					\item If Condition~D2 holds, there is a path from $\Omega_2$ to $\Omega_1$
					which may or may not go through~$m$.  
					\item If Condition~D3 holds, there is a node~$\tilde{m}$ from which there are paths to~$\Omega_1$ (which may or may not go through~$m$) 
					and to~$\Omega_2$.
				\end{itemize}
				Hence,  when Condition~C2 holds, there exists a path of the claimed form.

				\item The proof for Condition~C3 is the same as for Condition~C2.
			\end{itemize} 
			
			To finish the proof, we now show if there exists a path of the claimed form, then at least one of Conditions C1--C3 holds.     
			\begin{itemize}
				\item  If there is a single-edge path between~$\Omega_1$ and~$\Omega_2$, then Condition~C1 holds.
				
				\item Assume that there is a path with $L\ge 2$ edges from 
				$\Omega_1$~to~$\Omega_2$.  Denote by~$m$ the next to last node in the path. Then 
				$m \not\in \Omega_1 \cup \Omega_2$, and 
				the path without the
				last edge is a path from~$\Omega_1$ to~$\Omega_2 \cup \{m\}$. 
				Since $|\Omega_1 \cup \Omega_2 \cup \{m\}| =|\Omega_1 \cup \Omega_2|+1$,   then by the induction assumption, $Q_{\Omega_1,\Omega_2 \cup \{m\}}>0$. 
				In addition,$q_{m,k}>0$ for some $k\in \Omega_2$. Therefore, Condition~C3 holds.
				Similarly, if there is a path with $L>1$ edges from 
				$\Omega_2$~to~$\Omega_1$, then  Condition~C2 holds.
				
				\item  Finally, suppose that is some node~$\tilde{m} \notin \Omega_1\cup \Omega_2$ and paths from~$\tilde{m}$ to~$\Omega_1$ and to~$\Omega_2$. Since the
				case of a path from~$\Omega_1$ to~$\Omega_2$ or from~$\Omega_2$ to~$\Omega_1$ 
				has already been considered, we may assume that the path from~$\tilde{m}$ to~$\Omega_1$ contains no element of~$\Omega_2$ and vice versa. Also, by truncating the paths at the first node reached of the desired set, we may assume that no node of either
				path except the last belongs to~$\Omega_1 \cup \Omega_2$. Let $m$ be the next to last node of the path to~$\Omega_1$; note that $m$ might be~$\tilde{m}$.
				Since the path continues from~$m$ to~$\Omega_1$, then $q_{m,j}>0$ for some $j\in \Omega_1$.
				Moreover, the path from~$\tilde{m}$ to~$m$ is a path from~$\tilde{m}$ to~$\Omega_1\cup \{m\}$, so there exist paths from~$\tilde{m}$ to~$\Omega_1\cup \{m\}$ and from $\tilde{m}$ to~$\Omega_2$. Therefore, by the induction hypothesis, $Q_{\Omega_1\cup \{m\},\Omega_2}>0$. Hence, condition~C2 holds. 
			\end{itemize}
		\end{proof}

\begin{proof}[Proof of Theorem~{\rm \ref{thm:[Si,Sj]>[Si][Sj]-K}}]

	We proceed by induction on~$L$. The induction base $L=2$ is Lemma~\ref{lem:[Si,Sj]>[Si][Sj]}. 
		Assume that Theorem~{\rm \ref{thm:[Si,Sj]>[Si][Sj]-K}} holds for~$L$.
		To prove Theorem~{\rm \ref{thm:[Si,Sj]>[Si][Sj]-K}}  for~$L+1$,
		let us denote $\widetilde\Omega_1:=\bigcup_{l=1}^L \Omega_l$ and 
		$\widetilde\Omega_2:= \Omega_{L+1}$.
	Then
			\begin{equation*}
			[S_{\cup_{l=1}^{L+1}\Omega_l}]   = [S_{\widetilde\Omega_1,\widetilde\Omega_2}]   \ge 
			[S_{\widetilde\Omega_1}] \,[S_{\widetilde\Omega_2}]
			=[S_{\cup_{l=1}^{L}\Omega_l}] \, [S_{\Omega_{L+1}}] \ge \prod_{l=1}^{L+1} [S_{\Omega_l}] ,
				\end{equation*}
	   where the first inequality follows from Lemma~\ref{lem:[Si,Sj]>[Si][Sj]} and the second from the induction assumption. By Lemma~\ref{lem:[Si,Sj]>[Si][Sj]}, the first inequality is an equality
	   if and only there is no node which is influential 
		to both $\widetilde\Omega_1$ and~$\widetilde\Omega_2$.
		By the induction assumption, the second inequality is an equality if 
		and only if for any $l,  \widetilde{l} \in \{1, \dots, L\}$  where
		$l \not = \widetilde{l}$, 
		there is no node in~${\cal M}$ which is influential to both~$\Omega_l$ and~$\Omega_{\widetilde{l}}$. 
		Therefore, Theorem~{\rm \ref{thm:[Si,Sj]>[Si][Sj]-K}} follows for~$L+1$.
\end{proof}

		
		\subsection{Proof of Theorem~\ref{thm:[S_Omega]-prod[S_i]-N-paths}}
		\label{sec:[S_Omega]-prod[S_i]-N-paths}

		We first prove Theorem~\ref{thm:[S_Omega]-prod[S_i]-N-paths} for two sets that are connected by a {\em single path}:
		\begin{lemma}
			\label{lem:[S_i,j]-[S_i][S_j]-single-path}
			Consider the Bass/{\rm SI} model~\eqref{eqs:Bass-SI-models-ME} 
			on an undirected network, such that 
			\eqref{eqs:assumptions-upper-bound-SiSj} holds. 
			Let $\emptyset  \not= \Omega_1,\Omega_2 \subset {\cal M}$, such that
			$\Omega_1 \cap \Omega_2  = \emptyset$. 
			If there is a unique simple path~$\Gamma$ between~$\Omega_1$ and~$\Omega_2$, 
			then
			\begin{equation}
				\label{eq:[S_i,j]-[S_i][S_j]-single-path}
				[S_{\Omega_1,\Omega_2}]-[S_{\Omega_1}]  \, [S_{\Omega_2}]
				<
				E(t;K), \qquad t > 0 , 
			\end{equation}
			where  $K$~is the number of nodes of thge path~$\Gamma$, and
			$E(t;K)$ satisfies the bounds~\eqref{eq:[S_i,j]-[S_i][S_j]-N-paths-global-in-time-E-temporal} and~\eqref{eq:[S_i,j]-[S_i][S_j]-N-paths-global-in-time-E-global}.
			%
		\end{lemma} 
		\begin{proof}
			Let $t>0$. 
			%
			Denote by~${\cal N}^-$ the network obtained by deleting the {\em central edge} of~$\Gamma$.\footnote{If $K$ is even, we delete the $\frac{K}{2}$th edge. If $K$ is odd, we delete either the $\frac{K-1}{2}$th or the $\left(\frac{K-1}{2}+1\right)$th edge} In this network, there is no node which is influential 
			to~$\Omega_1$ and to~$\Omega_2$, see Corollary~\ref{cor:paths-influential-node}. Hence, by Lemma~\ref{lem:[Si,Sj]>[Si][Sj]},
			$$
			[S_{\Omega_1,\Omega_2}^-] = [S_{\Omega_1}^-] \,[S_{\Omega_2}^-],
			$$ 
			where $[S_{\Omega}^-]$ denotes the nonadoption probability of~$\Omega$  in~${\cal N}^-$. Since the deleted edge is influential to~$\Omega_1$ and~$\Omega_2$, 
			it follows from the {\em dominance principle}, see~\cite{Bass-boundary-18}, that 
			$$
			[S_{\Omega_1}] < [S_{\Omega_1}^-], \qquad
			[S_{\Omega_2}] < [S_{\Omega_2}^-], \qquad
			[S_{\Omega_1,\Omega_2}] < [S_{\Omega_1,\Omega_2}^-].
			$$
			Therefore,
			\begin{equation*}
				\begin{aligned}
					[S_{\Omega_1,\Omega_2}]-[S_{\Omega_1}]  \, [S_{\Omega_2}]& 
					< [S_{\Omega_1,\Omega_2}^-] - [S_{\Omega_1}]  \, [S_{\Omega_2}] 
					 = [S_{\Omega_1,\Omega_2}^-] - [S_{\Omega_1}^-] \,[S_{\Omega_2}^-]+[S_{\Omega_1}^-] \,[S_{\Omega_2}^-] - [S_{\Omega_1}]  \, [S_{\Omega_2}] 
					\\ &  =
					[S_{\Omega_1}^-] \,[S_{\Omega_2}^-] - [S_{\Omega_1}]  \, [S_{\Omega_2}] 
					= ([S_{\Omega_1}^-]  - [S_{\Omega_1}])\,[S_{\Omega_2}^-] +  ([S_{\Omega_2}^-]  - [S_{\Omega_2}])\,[{S_{\Omega_1}}]. 
				\end{aligned}
			\end{equation*}
			Since $0< [S_{\Omega_2}^-], [{S_{\Omega_1}}] <1$, see~\eqref{eq:[S]<e^-pt}, we have that 
			\begin{subequations}
				\label{eqs:[S_i,j]-[S_i][S_j]-single-path-interim-inequalities}
				\begin{equation}
					\label{eq:[S_i,j]-[S_i][S_j]-derivation}
					[S_{\Omega_1,\Omega_2}]-[S_{\Omega_1}]  \, [S_{\Omega_2}] < \Big([S_{\Omega_1}^-]  - [S_{\Omega_1}]\Big) + \Big([S_{\Omega_2}^-]  - [S_{\Omega_2}]\Big).
				\end{equation}
				
				Denote by~$m_1$ and~$m_2$ the nodes of the deleted central edge which are connected  in~${\cal N}^-$ to~$\Omega_1$ and to~$\Omega_2$, respectively, and by~${\cal N}^+$ the network that is obtained by transferring the two directional weights of the deleted edge to the nodes~$m_1$ and~$m_2$, i.e., by setting 
				$$
				q_{m_2,m_1}^+=q_{m_1,m_2}^+=0, \qquad p_{m_1}^+=p_{m_2}^+=p+q.
				$$
				Denote the probabilities in~${\cal N}^+$ by~$[S_{\Omega}^+]$. 
				By the dominance principle,
				$[S_{\Omega_1}] >[S_{\Omega_1}^+]$ and $[S_{\Omega_2}] >[S_{\Omega_2}^+]$. 
				Hence,
				\begin{equation}
					\label{eq:[S_i,j]-[S_i][S_j]-single-path-interim-inequalities-b}
					[S_{\Omega_1}^-]  - [S_{\Omega_1}]<[S_{\Omega_1}^-]  - [S_{\Omega_1}^+], \qquad
					[S_{\Omega_2}^-]  - [S_j]<[S_{\Omega_2}^-]  - [S_{\Omega_2}^+].
				\end{equation}
				Combining inequalities~\eqref{eqs:[S_i,j]-[S_i][S_j]-single-path-interim-inequalities}, 
				we obtain~\eqref{eq:[S_i,j]-[S_i][S_j]-single-path} with 
				$$
				E(t;K):= \Big( [S_{\Omega_1}^-]  - [S_{\Omega_1}^{+}]\Big)+\Big( [S_{\Omega_2}^-]  - [S_{\Omega_2}^+]\Big).
				$$
				
				Next, we derive the bound~\eqref{eq:[S_i,j]-[S_i][S_j]-N-paths-global-in-time-E-temporal} for~$E(t;K)$.
				Denote by $i_1 \in \Omega_1$ and $i_2 \in \Omega_2$ the end nodes of the path~$\Gamma$.  
				The difference  $[S_{\Omega_1}^-] - [S_{\Omega_1}^+]$ is only due to realizations in which 
				$i_1$~adopts because of an {\em adoption path}
				from~$m_1$ to~$\Omega_1$ in~${\cal N}^+$, but not in~${\cal N}^-$.
				Therefore,
				$$
				[S_{\Omega_1}^-]  - [S_{\Omega_1}^{+}]<[S_{i_1}^-] - [S_{i_1}^+].
				$$
				Denote by $\widetilde{\cal N}^+$ and~$\widetilde{\cal N}^-$ the networks obtained from 
				${\cal N}^+$ and~${\cal N}^-$ by keeping only the nodes and edges from~$m_1$ to~$i_1$,
				and denote  the probabilities in~$\widetilde{\cal N}^\pm$ by~$[\widetilde S_{\Omega}^\pm]$.
				Then 
				$$
		[S_{i_1}^-] - [S_{i_1}^+] <  [\widetilde S_{i_1}^-] - [\widetilde S_{i_1}^+].
		$$
%
%
				The networks $\widetilde{\cal N}^+$ and~$\widetilde{\cal N}^-$ 
			are the homogeneous and heterogeneous one-sided lines with $p_1=p$ and $\bar{p}_1:= p+q$
				that were defined in~\cite[Lemma~15]{OR-24},
				and 
				the number of nodes in these lines is either
				$\left \lfloor{\frac{K}{2}}\right \rfloor $
				or~$\left \lfloor{\frac{K}{2}}\right \rfloor +1$. 
				Therefore,   by~\cite[eq.~(34)]{OR-24},
				\begin{equation}
					\label{eq:[S_i^-]-[S_i^+]<}
					[S_{\Omega_1}^-]  - [S_{\Omega_1}^+]
					< [S_{i_1}^-] - [S_{i_1}^+]
					<[\widetilde S_{i_1}^-] - [\widetilde S_{i_1}^+]
					 <
					  (1-I^0)e^{-\left(p+q\right)t}\left(\frac{eqt}{\left \lfloor{\frac{K}{2}}\right \rfloor}\right)^{\left \lfloor{\frac{K}{2}}\right \rfloor},
					\quad  \left \lfloor{\frac{K}{2}}\right \rfloor >  qt.
				\end{equation}
			\end{subequations}
			The same bound also holds for~$[S_{\Omega_2}^-]  - [S_{\Omega_2}^+]$.
			Therefore, we obtain~\eqref{eq:[S_i,j]-[S_i][S_j]-N-paths-global-in-time-E-temporal}. 
			
			Finally, to prove the globally-uniform upper bound~\eqref{eq:[S_i,j]-[S_i][S_j]-N-paths-global-in-time-E-global}, 
			we note that, by Lemma~\ref{lem:[S_Omega]<=[S_Omega^0]e^-pt}, 
			\begin{equation}
				\label{eq:[S_i^-]-[S_i^+]<e^-pt}
				[S_{\Omega_1}^-]  - [S_{\Omega_1}^+]< (1-I^0)e^{-pt},
				\qquad   t > 0 .
			\end{equation}
			As in the proof of~\cite[Corollary~3]{OR-24},
			from inequalities~\eqref{eq:[S_i^-]-[S_i^+]<} and~\eqref{eq:[S_i^-]-[S_i^+]<e^-pt} it follows that
			\begin{equation}
				\label{eq:[S_i^-]-[S_i^+]-global}
				[S_{\Omega_1}^-]  - [S_{\Omega_1}^+] 
				<
				(1-I^0)  \left(\frac{q}{p+q} \right)^{\left \lfloor{\frac{K}{2}}\right \rfloor},
				\qquad  t > 0 .
			\end{equation}
			The same bound also holds for~$[S_{\Omega_2}^-]  - [S_{\Omega_2}^+]$. Therefore,  
			we have~\eqref{eq:[S_i,j]-[S_i][S_j]-N-paths-global-in-time-E-global}.
			%
			%
			%
			%
			%
		\end{proof}

		Next, we consider two sets that are connected by $N$~paths:
		\begin{lemma}
			\label{lem:[S_i,j]-[S_i][S_j]-N-paths}
			Consider the Bass/{\rm SI} model~\eqref{eqs:Bass-SI-models-ME} 
			on an undirected network, such that 
			\eqref{eqs:assumptions-upper-bound-SiSj} holds. 
			Let $\emptyset \not= \Omega_1,\Omega_2 \subset {\cal M}$, such that
			$\Omega_1 \cap \Omega_2  = \emptyset$. 
			If there are $N \ge 2$ distinct simple paths $\{\Gamma_n\}_{n=1}^N$ between~$\Omega_1$ and~$\Omega_2$, such that their interior nodes are in~${\cal M} \setminus \left(\Omega_1 \cup \Omega_2\right)$, then
			\begin{equation}
				\label{eq:[S_i,j]-[S_i][S_j]-N-paths}
				[S_{\Omega_1,\Omega_2}]-[S_{\Omega_1}]  \, [S_{\Omega_2}]< 
				\sum_{n=1}^N 
				E(t;K_n), \qquad t > 0  ,	
			\end{equation}
			where $K_n$ is the number of nodes of the path~$\Gamma_n$, 
			and~$E(t;K_n)$ satisfies the bounds~\eqref{eq:[S_i,j]-[S_i][S_j]-N-paths-global-in-time-E-temporal} and~\eqref{eq:[S_i,j]-[S_i][S_j]-N-paths-global-in-time-E-global}.
			%
			%
		\end{lemma}
		\begin{proof}
			Denote the end nodes of the path~$\Gamma_n$ 
			by~$i_{1,n} \in \Omega_1 $ and~$i_{2,n} \in \Omega_2$.  
			{\em Assume first that the $N$ paths are disjoint}, i.e., that do not share interior nodes (they may share, however,
			the end nodes $\{i_{1,n}\}$ and~$\{i_{2,n}\}$).
			Denote by~${\cal N}^-$ the network obtained by deleting the $N$ 
			central edges of~$\{\Gamma_n\}_{n=1}^N$. Then, as in the proof of 
			Lemma~\ref{lem:[S_i,j]-[S_i][S_j]-single-path}, see~\eqref{eq:[S_i,j]-[S_i][S_j]-derivation}, 
			\begin{subequations}
				\label{eqs:[S_i,j]-[S_i][S_j]-N-paths-interim-inequalities}
				\begin{equation}
				\label{eq:[S_i,j]-[S_i][S_j]-N-paths-interim-inequalities-a}
					[S_{\Omega_1,\Omega_2}]-[S_{\Omega_1}]  \, [S_{\Omega_2}] <   \Big([S_{\Omega_1}^-]  - [S_{\Omega_1}]\Big) +  \Big([S_{\Omega_2}^-]  - [S_{\Omega_2}]\Big).
				\end{equation}
				For $n = 1, \dots, N$,
				denote by $m_{1,n}$ and~$m_{2,n}$ the nodes of the deleted central edge 
				of~$\Gamma_n$, which are connected in~${\cal N}^-$ to~$\Omega_1$
				and to~$\Omega_2$, respectively. Denote by~${\cal N}^+$ the network obtained by transferring the  $2n$ directional weights of the deleted  edges to the $2n$ nodes
				of these edges, i.e., 
				$$
				q_{m_{2,n},m_{1,n}}^+=q_{m_{1,n},m_{2,n}}^+=0, \quad p_{m_{1,n}}^+=p_{m_{2,n}}^+=p+q, \qquad 
				n = 1, \dots, N.
				$$
				
				As in the proof of Lemma~\ref{lem:[S_i,j]-[S_i][S_j]-single-path},
				see~\eqref{eq:[S_i,j]-[S_i][S_j]-single-path-interim-inequalities-b}, 
				\begin{equation}
					\label{eq:[S_i,j]-[S_i][S_j]-N-paths-interim-inequalities-b}
					[S_{\Omega_1}^-]  - [S_{\Omega_1}]<[S_{\Omega_1}^-]  - [S_{\Omega_1}^+].
				\end{equation}
				The difference between $[S_{\Omega_1}^-]$ and~$[S_{\Omega_1}^+]$ is due to realizations in which 
				$\Omega_1$~adopts because of one of the $N$ adoption paths from~$m_{1,n}$ to~$\Omega_1$ in~${\cal N}^+$, but not in~${\cal N}^-$.
				Therefore, it is bounded by the sum of the individual differences in~$[S_{\Omega_1}]$  due to each of these $N$~paths, i.e.,
\begin{equation}
	\label{eq:[S_i,j]-[S_i][S_j]-N-paths-interim-inequalities-c}
		[S_{\Omega_1}^-]  - [S_{\Omega_1}^+] \le \sum_{n=1}^N \Big( [S_{\Omega_1}^-]  - [S_{\Omega_1}^{+,n}]\Big),
\end{equation}
\end{subequations} 
where~$[S_{\Omega_1}^{+,n}]$ is the nonadoption probability of~$\Omega_1$ in the network~${\cal N}^{+,n}$, that is obtained from~${\cal N}^{-}$ by 
			setting 
			\begin{equation}
				\label{eq:N^+,n-N-paths}
	q_{m_{2,n},m_{1,n}}^+=q_{m_{1,n},m_{2,n}}^+=0, \qquad 				p_{m_{1,n}}^{+,n}=p_{m_{2,n}}^{+,n}=p+q.
			\end{equation}
			Combining inequalities~\eqref{eqs:[S_i,j]-[S_i][S_j]-N-paths-interim-inequalities}, 
			and noting that~\eqref{eq:[S_i,j]-[S_i][S_j]-N-paths-interim-inequalities-b}
			and~\eqref{eq:[S_i,j]-[S_i][S_j]-N-paths-interim-inequalities-c}
			also hold for node~$\Omega_2$,  we 
			obtain 
			\begin{equation}
			\label{eq:[S_i,j]-[S_i][S_j]-N-paths-in-proof}
			[S_{\Omega_1,\Omega_2}]-[S_{\Omega_1}]  \, [S_{\Omega_2}]< 
			\sum_{n=1}^N 
			E(t;K_n), \quad
			E(t;K_n):= \Big( [S_{\Omega_1}^-]  - [S_{\Omega_1}^{+,n}]\Big)+\Big( [S_{\Omega_2}^-]  - [S_{\Omega_2}^{+,n}]\Big).
			\end{equation}

			%
			%

			{\em Let us now derive the bounds}~\eqref{eq:[S_i,j]-[S_i][S_j]-N-paths-global-in-time-E-temporal}
			and~\eqref{eq:[S_i,j]-[S_i][S_j]-N-paths-global-in-time-E-global}  for~$E(t;K_n)$.
			By~\eqref{eq:[S_i^-]-[S_i^+]<} and~\eqref{eq:N^+,n-N-paths},  
			\begin{equation}
				\label{eq:[S_i,j]-[S_i][S_j]-N-paths-interim-inequalities-d}
				[S_{\Omega_1}^-]  - [S_{\Omega_1}^{+,n}]
				<
				(1-I^0)
				e^{-\left(p+q\right)t}\left(\frac{eqt}{\left \lfloor{\frac{K_n}{2}}\right \rfloor}\right)^{\left \lfloor{\frac{K_n}{2}}\right \rfloor},
				\qquad  \left \lfloor{\frac{K_n}{2}}\right \rfloor >   qt.
			\end{equation}
			The same bound  also hold for~$[S_{\Omega_2}^-]  - [S_{\Omega_2}^{+,n}]$. Therefore, we obtain~\eqref{eq:[S_i,j]-[S_i][S_j]-N-paths-global-in-time-E-temporal}. 
						Finally, as in the proof of Lemma~\ref{lem:[S_i,j]-[S_i][S_j]-single-path}, for all $t>0$ we have that
			$$
			[S_{\Omega_1}^-]  - [S_{\Omega_1}^{+,n}]< (1-I^0)e^{-pt}, \qquad t>0. 
			$$
			From this inequality and~\eqref{eq:[S_i,j]-[S_i][S_j]-N-paths-interim-inequalities-d} it follows that
			$$	
			[S_{\Omega_1}^-]  - [S_{\Omega_1}^{+,n}] 
			<
			(1-I^0)
			\left(\frac{q}{p+q} \right)^{\left \lfloor{\frac{K_n}{2}}\right \rfloor},
			\qquad   t>0.
			$$
			The same bound also holds for~$[S_{\Omega_2}^-]  - [S_{\Omega_2}^{+,n}]$. Hence, we obtain~\eqref{eq:[S_i,j]-[S_i][S_j]-N-paths-global-in-time-E-global}.

{\em Consider now the case where the paths  $\{\Gamma_n\}_{n=1}^N$  are not disjoint}. 
Note that in this case,  $\{\Gamma_n\}_{n=1}^N$  refers to all the possible paths between~$\Omega_1$ 
	and~$\Omega_2$. Thus, for example, two paths that intersect at a single node  are counted as four different paths. Similarly, if two paths merge into a single path, then separate, then merge into a single path, they are also counted as four paths.
Without loss of generality, we can assume that the paths are arranged in order of increasing length, so that $K_1 \le K_2 \cdots \le K_N$.  
We construct the network~${\cal N}^-$ iteratively, as follows. For $n=1, \dots, N$, if after the $n-1$th iteration all the edges of the path~$\Gamma_n$ still exist, we 
delete the central edge of~$\Gamma_n$. 
At the end of this iterative process,  
all the $N$ paths are disconnected, and so the sets $\Omega_1$ and $\Omega_2$ are disjoint in~${\cal N}^-$. 
Therefore, $[S_{\Omega_1,\Omega_2}^-] = [S_{\Omega_1}^-] \,[S_{\Omega_2}^-]$,
and so \eqref{eq:[S_i,j]-[S_i][S_j]-N-paths-interim-inequalities-a} holds.
As before, let~${\cal N}^+$ be obtained from~${\cal N}^-$ by increasing the weights of the nodes of the deleted edges 
from~$p$ to~$p+q$. Then   \eqref{eq:[S_i,j]-[S_i][S_j]-N-paths-interim-inequalities-b} holds. 
We claim that the bound~\eqref{eq:[S_i,j]-[S_i][S_j]-N-paths-interim-inequalities-c} also holds, and therefore that~\eqref{eq:[S_i,j]-[S_i][S_j]-N-paths-in-proof} holds. 
			  
 To prove that 
 \eqref{eq:[S_i,j]-[S_i][S_j]-N-paths-interim-inequalities-c} still~holds, 
 we first note that, as in the case  of disjoint paths, the difference $[S_{\Omega_1}^-]  - [S_{\Omega_1}^{+}]$ 
is only due to the adoption paths that start from the nodes of the deleted edges
and reach $\Omega_1$ in~${\cal N}^-$ (and in~${\cal N}^+$).
Unlike the case of disjoint paths, however, in the networks~${\cal N}^-$ and~${\cal N}^+$, there can be more than one adoption path 
from a node of a deleted edge to~$\Omega_1$. Moreover, these adoption paths can intersect or even share edges. 
Nevertheless, since the overall contribution to $[S_{\Omega_1}^-]  - [S_{\Omega_1}^{+}]$ from these adoption paths 
 is due to realizations in which 
 $\Omega_1$~adopts because of one of these adoption paths in~${\cal N}^+$ but not in~${\cal N}^-$,
 it is still bounded by the contribution due to each of these adoption paths separately.
 Therefore, we now consider the separate contribution of each of these adoption paths.

 Assume that in the $n$th iteration  in the construction of~${\cal N}^-$, 
 we deleted the central edge 
 $m_{1,n} \leftrightarrow m_{2,n}$ of the path~$\Gamma_n$.
 Denote by $\Gamma_n^1$ and $\Gamma_n^2$ the equal-length subpaths of~$\Gamma_n$ between 
$\Omega_1$ and~$m_{1,n}$ and between $m_{2,n}$ and~$\Omega_2$, respectively, i.e.,
$$
			\Gamma_n =  \Gamma_n^1 \leftrightarrow m_{1,n}\leftrightarrow m_{2,n}\leftrightarrow \Gamma_n^2.
$$
In the network~${\cal N}^+$, the node $m_{1,n}$ has weight $p+q$. 
 The contribution to the difference $[S_{\Omega_1}^-]  - [S_{\Omega_1}^+]$ 
of the adoption path from~$m_{1,n}$ to~$\Omega_1$ through $\Gamma_n^1$  is bounded by the $n$th~term in the sum~\eqref{eq:[S_i,j]-[S_i][S_j]-N-paths-interim-inequalities-c}.  
We also need to consider, however, the possibility that in the network~${\cal N}^+$, 
 the node  $m_{1,n}$ is connected to~$\Omega_1$  through another subpath,
 which we denote by $\widetilde{\Gamma}^1$.  Let us also denote the path
 between $\Omega_1$ and $\Omega_2$ which is made of $\widetilde{\Gamma}^1$
 and $\Gamma_n^2$ by $\Gamma_{\tilde{n}}$, i.e., 
  $$
 \Gamma_{\tilde{n}}:=\widetilde{\Gamma}^1 \leftrightarrow m_{1,n}\leftrightarrow m_{2,n}\leftrightarrow \Gamma_n^2.
 $$
\begin{itemize}
  	\item If $\widetilde{\Gamma}^1$ is shorter than  $\Gamma^1_n$, the  path $\Gamma_{\tilde{n}}$
    is shorter than~$\Gamma_n$. Since the subpath 
   $\widetilde{\Gamma}^1$ exists in~${\cal N}^+$, the path~$\Gamma_{\tilde{n}}$ exists at the beginning of the $n$th iteration. This, however, is in contradiction with the iterative construction of~${\cal N}^-$, since $\Gamma_{\tilde{n}}$ is shorter than  $\Gamma_n$.

	\item If $\widetilde{\Gamma}^1$ is longer than  $\Gamma^1_n$, 
 	the  path $\Gamma_{\tilde{n}}$
 	is longer than~$\Gamma_n$, and so $\tilde n>n$. At the $\tilde n$th iteration, the path $\Gamma_{\tilde{n}}$ does not exist (since we 
 	already deleted the edge $m_{1,n}\leftrightarrow m_{2,n}$). 
 	In the sum~\eqref{eq:[S_i,j]-[S_i][S_j]-N-paths-interim-inequalities-c},
 	however, we accounted for the impact of deleting the central edge of 
 	$\Gamma_{\tilde{n}}$ by the term with~$K_{\tilde{n}}$.  This term is  
 	larger than the one needed for the impact of 
 	the node  $m_{1,n}$ on~$[S_{\Omega_1}^-]  - [S_{\Omega_1}^+]$  
 	through~$\widetilde{\Gamma}^1$, since $\widetilde{\Gamma}^1$ is longer than  $\Gamma^2_n$, and so the central edge of  $ \Gamma_{\tilde{n}}$ lies inside $\Gamma^1_n$ (i.e., is closer to $\Omega_1$ than $m_{1,n}$).

   	\item If $\widetilde{\Gamma}^1$ has the same length as $\Gamma^1_n$,
   	then we can assume without loss of generality that $\hat n>n$,
   	 and therefore a similar argument holds.

%
%
	  \end{itemize}

Finally, we need to rule out the possibility that in the network~${\cal N}^+$ (in which the edge  $m_{1,n} \leftrightarrow m_{2,n}$ has been deleted), 
the node  $m_{2,n}$ is also connected to~$\Omega_1$.
Indeed, assume by contradiction that $m_{2,n}$ is connected to~$\Omega_1$ in~${\cal N}^+$. 
Since there is no path between $\Omega_1$ and $\Omega_2$ in~${\cal N}^+$,  this implies that
there is no path  between~$m_{2,n}$ and~$\Omega_2$ in~${\cal N}^+$.
Since, however, the path $\Gamma_n^2$ between~$m_{2,n}$ and~$\Omega_2$ exists at the end of the $n$th~iteration, 
this implies that at some later iteration $\hat{n} > n$, the path~$\Gamma_n^2$ became disconnected because one of the edges of was deleted.
This deleted edge is the central edge $m_{1,\hat n}\leftrightarrow m_{2, \hat n}$ of the path
  $$
\Gamma_{\hat{n}}:={\Gamma}^1_{\hat n} \leftrightarrow m_{1,\hat n}\leftrightarrow m_{2, \hat n}\leftrightarrow \Gamma_{\hat n}^2.
$$
Since the central edge of~$\Gamma_{\hat{n}}$ was deleted
at the $\hat{n}$th iteration, the path~$\Gamma_{\hat n}$ existed at the 
beginning of the $\hat{n}$th iteration. 
Let $\bar{\Gamma}^2$ denote the subpath of $\Gamma_n^2$ between~$ m_{2, \hat n}$ and~$\Omega_2$. Then  at the 
beginning of the $\hat{n}$th iteration, 
the path 
  $$
\bar{\Gamma}:={\Gamma}^1_{\hat n} \leftrightarrow m_{1,\hat n} \leftrightarrow m_{2, \hat n}\leftrightarrow \bar{\Gamma}^2
$$
between $\Omega_1$ and $\Omega_2$ also exists. The length of~$\bar{\Gamma}$, however, is shorter than that of $\Gamma_{\hat{n}}$.\,\footnote{Since $\bar{\Gamma}^2$ is a proper subpath of~$\Gamma_n^2$, it is shorter than $\Gamma_n^2$, which in turn is shorter than $\Gamma_{\bar n}^2$ (since $n< \bar n$).}
 Therefore, we reached a 
contradiction, since the central edge of
$\bar{\Gamma}$ should have been deleted before that of~$\Gamma_{\hat{n}}$,
and so $\bar{\Gamma}$ could not exist  at the 
beginning of the $\hat{n}$th iteration.
		\end{proof}

			\begin{proof}[Proof of Theorem~{\rm \ref{thm:[S_Omega]-prod[S_i]-N-paths}}]
			
			   When $N_L \ge 1$, it follows from 
			   Lemma~\ref{lem:paths-influential-node} that there exists a node which is influential to both $\Omega_1$ and $\Omega_2$. 
			   Therefore, the lower bound follows from Theorem~\ref{thm:[Si,Sj]>[Si][Sj]-K}.
			   
				For the upper bound, we proceed by induction on~$L$. The case $L=2$ is Lemma~\ref{lem:[S_i,j]-[S_i][S_j]-N-paths}. 
				Assume that~\eqref{eq:[S_Omega]-prod[S_i]-N-paths} holds for~$L$.
				Consider~\eqref{eq:[S_Omega]-prod[S_i]-N-paths}  for~$L+1$.
				We can reorder the~$N_{L+1}$ paths among~$\{\Omega_1, \dots, \Omega_{L+1} \}$,
				so that the first $N_L$ paths are among~$\{\Omega_1, \dots, \Omega_L \}$,
				and the paths between~$\{\Omega_1, \dots, \Omega_L \}$ and~$\Omega_{L+1}$ are enumerated 
				from~$N_L+1$ to~$N_{L+1}$.  
							Therefore, by~\eqref{eq:[S_i,j]-[S_i][S_j]-N-paths}
				with $\widetilde\Omega_1:=\bigcup_{l=1}^L \Omega_l$ and 
				$\widetilde\Omega_2:= \Omega_{L+1}$,  
				$$
				[S_{\Omega_1, \dots, \Omega_{L+1}}]-[S_{\Omega_1, \dots, \Omega_L}] \, [S_{\Omega_{L+1}}]
				=	[S_{\widetilde\Omega_1,\widetilde\Omega_2}]-[S_{\widetilde\Omega_1}]  \, [S_{\widetilde\Omega_2}]
				<
				\sum_{n=N_L+1}^{N_{L+1}} 
				E(t;K_n).
				$$
				Hence, since $ [S_{\Omega_{L+1}}] \le 1$, 
				$$
				\begin{aligned}
					[S_{\Omega_1, \dots, \Omega_{L+1}}]-\prod_{l=1}^{L+1} [S_{\Omega_l}]
					& = 
					\Big(	[S_{\Omega_1, \dots, \Omega_{L+1}}]- [S_{\Omega_1, \dots, \Omega_L}]  [S_{\Omega_{L+1}}]\Big)
					+ \Big(	 [S_{\Omega_1, \dots, \Omega_L}]  [S_{\Omega_{L+1}}] -\prod_{l=1}^{L+1} [S_{\Omega_l}]\Big)
					\\ & <
					\sum_{n=N_L+1}^{N_{L+1}} 
					E(t;K_n)
					+ 
					[S_{\Omega_{L+1}}] \Big(	 [S_{\Omega_1, \dots, \Omega_L}]  -\prod_{l=1}^{d} [S_{\Omega_l}]\Big)
					\\ & <
					\sum_{n=N_L+1}^{N_{L+1}}
					E(t;K_n)
					+ 
					\sum_{n=1}^{N_L}
					E(t;K_n).
				\end{aligned}
				$$ 
				Therefore, we have~\eqref{eq:[S_Omega]-prod[S_i]-N-paths}. 
			\end{proof}

			\section{Proving the funnel theorems}
			\label{sec:ptoofs-funnel}

			The adoption/infection of node~$j$ in network~${\cal N}$
			is due to one of the following $L+1$ distinct influences:
			\begin{enumerate}
				\item Internal influences on~$j$ by edges that arrive from~$A_l$ for some $l \in \{1, \dots, L\}$. 
				\item External influences on~$j$.
			\end{enumerate} 
			In order to identify the specific influence that leads to the adoption of~$j$, we introduce
			\begin{definition} [Network~${\cal N}^{A_l}$]
				\label{def:networks-N^A,N^B}
				Consider the Bass/{\rm SI}  model~\eqref{eqs:Bass-SI-models-ME} on 
				network~${\cal N}$.  
				Let $j \in \cal M$ and $A_l \subset {\cal M}$.
				The network ${\cal N}^{A_l}$ is obtained from~${\cal N}$ by removing all influences on node~$j$, except for directed edges from~$A_l$ to~$j$. Thus, we set~$p_j:=0$,   $I_j^0:=0$, and
				$q_{k,j} := 0$ for~$k\in {\cal M} \setminus A_l $. 
				\end{definition}

				The {\em funnel inequality} shows that the nonadoption probability 
				$
				[S_{{j}}]
				$
				is bounded from below by the product of the nonadoption probabilities of~$j$ due to each of the $L+1$ distinct influences:
				\begin{theorem}
			\label{thm:funnel_node_inequality}
			Consider the Bass/{\rm SI}  model~\eqref{eqs:Bass-SI-models-ME}.
			Let $\{A_1, \dots, A_L,\{j\}\}$ be a partition of~$ {\cal M}$.
			Then
				\begin{equation}
					\label{eq:funnel_inequality}
					[S_{{j}}]  \ge
					[S^{p_j}_j] \, 	\prod_{l=1}^L [S^{A_l}_j] ,
					\qquad t \ge 0,  
				\end{equation}
				where $[S^{p_j}_j] = (1-I^{0}_j) \, e^{-p_j t}$.  
			\end{theorem}
			\begin{proof}
				In network~${\cal N}^{p_j}$,  $j$ is an isolated node. Hence,  the expression for~$[S^{p_j}_j]$ 
				follows from the master equations~\eqref{eqs:master-eqs-general}.
				
				To prove~\eqref{eq:funnel_inequality}, we first note that by the indifference principle (Theorem~\ref{thm:Indifference}), all the edges that emanate from~$j$ are non-influential to~$j$. Since this holds for all the $L+2$~probabilities in~\eqref{eq:funnel_inequality}, in what follows, {\em we can assume that no edges emanate from~$j$}. 
				
				In principle, we need to compute the $L+2$~probabilities in~\eqref{eq:funnel_inequality} on the $L+2$~networks 
				${\cal N}$, ${\cal N}^{A_1}$, \dots, ${\cal N}^{A_L}$, and~${\cal N}^{p_j}$.
				We can simplify the analysis, however, by considering only two networks, as follows.  
				Given the original network~${\cal N}$, we define the network~${\cal N}^+$ by ``splitting'' node~$j$ into the $L+1$~nodes  $\{{j}_{A_1}, \dots, {j}_{A_L}, {j}_p\}$, such that:
				\begin{enumerate}
					\item ${j}_{A_l}$ inherits from~$j$  the directed edges from~$A_l$ to~$j$, i.e.,
					$$
					[S_{j_{A_l}}^{+}](0):=1, \qquad 
					{p}_{{j}_{A_l}}^+:=0, \quad 
					{q}_{k,{j}_{A_l}}^+:= {q}_{k,j}\, \mathbbm{1}_{k \in A_l }, 
					\quad  k \in {\cal M}, 
					\qquad i=1, \dots, K.
					$$
					\item  ${j}_p$ inherits from $j$ its weight and initial condition, 
					i.e.,
					$$
					[S_{j_p}^{+}](0):= [S_j^0], \qquad 
					{p}_{{j}_{p}}^+:=p_j, \qquad 
					{q}_{k,{j}_p}^+ := 0, \quad k \in \cal M.
					$$
					
					
					\item Since  no edges emanate from~$j$ in network~${\cal N}$, no edges emanate from~${j}_{A_1}$, \dots,  ${j}_{A_L}$, and~${j}_p$ in network~${\cal N}^+$.
					
					\item The weights of the nodes~${\cal M} \setminus \{j\}$, and of the edges among these nodes, are the same in~${\cal N}$ and in~${\cal N}^+$.
					
				\end{enumerate}
				
				Let ${X}^{+}_{k}(t)$ denote the state of node~$k$ in network ${\cal N}^+$,
				and let $[S^{+}_k]:=\mathbb{P}({X}^{+}_{k}(t)=0)$. 
				By construction, 
				\begin{equation}
					\label{eq:Prob(X^j,A)=Prob(X^j,tilde-A)}
					[S^{p_j}_j] =  [S^{+}_{{j}_p}], 
					\qquad 		 
					[S^{A_l}_j] = 
					[S^{+}_{{j}_{A_l}}],
					\quad l \in \{1, \dots, L\}.
				\end{equation}
				In Appendix~\ref{app:Prob(X^j,A)},  we prove that 
				\begin{equation}
					\label{eq:Prob(X^j,A)}
					[S_{{j}}] = 
					[S^{+}_{{j}_{A_1}, \dots, {j}_{A_L},{j}_p}],
				\end{equation}
				where 
				$[S^{+}_{{j}_{A_1}, \dots, {j}_{A_L},{j}_p}]:=
				\mathbb{P} \big({X}^{+}_{{j}_{A_1}}(t)
				= \dots = {X}^{+}_{{j}_{A_L}}(t) ={X}^{+}_{{j}_p}(t)
				=0\big)$.
				Since $ {j}_p$ is an isolated node in~${\cal N}^+$, its adoption is independent of that of $j_{A_1}, \dots,  {j}_{A_L}$, and so
				\begin{equation}
					\label{eq:Prob(X^j,A)-indifference}
					[S^{+}_{{j}_{A_1}, \dots, {j}_{A_L},{j}_p}] = 	[S^{+}_{{j}_p}] \, 	[S^{+}_{{j}_{A_1}, \dots, {j}_{A_L}}].
				\end{equation}
				Applying Theorem~\ref{thm:[Si,Sj]>[Si][Sj]-K} to network~${\cal N}^+$ gives
				\begin{equation}
					\label{eq:Prob(X^j,A)-conditional}
					[S^{+}_{{j}_{A_1}, \dots, {j}_{A_L}}]
					\ge   \prod_{l=1}^L	[S^{+}_{{j}_{A_l}}].
				\end{equation}
				Combining relations~\eqref{eq:Prob(X^j,A)}, 
				\eqref{eq:Prob(X^j,A)-indifference}, 
				and~\eqref{eq:Prob(X^j,A)-conditional}  gives
				\begin{equation}
					\label{eq:Prob(X^j,A)+indifference}
					[S_{{j}}]
					\ge
					[S^{+}_{{j}_p}] \,
					\prod_{l=1}^L	[S^{+}_{{j}_{A_l}}].
				\end{equation}
				Substituting~\eqref{eq:Prob(X^j,A)=Prob(X^j,tilde-A)} in~\eqref{eq:Prob(X^j,A)+indifference} proves~\eqref{eq:funnel_inequality}. 
			\end{proof}


			\begin{lemma}
			\label{lem:funnel_node_iequality}
			Consider the Bass/{\rm SI}  model~\eqref{eqs:Bass-SI-models-ME}.
			Let $j \in \cal M$, and let $\{A_1, \dots, A_L,\{j\}\}$ be a partition of~${\cal M}$. 
			\begin{itemize}
				\item 
				
				If 
				$j$ is a funnel node of~$\{A_1\}_{l=1}^L$,  then   
				\begin{equation}
					\label{eq:funnel_equality}
					[S_{{j}}] =
					[S^{p_j}_j] \, \prod_{l=1}^L [S^{A_l}_j],  \qquad t \ge0.  \qquad  \mbox{\bf (funnel  equality)}
				\end{equation}
				
				\item 
				If, however, 
				$j$ is not a funnel node of~$A_1$ and~$A_2$, then 
				\begin{equation}
					\label{eq:funnel_strong_jnequality}
					[S_{{j}}]  >
					[S^{p_j}_j] \, \prod_{l=1}^L [S^{A_l}_j],
					\qquad t>0.  \qquad  \mbox{\bf (strict funnel  inequality)}
				\end{equation}
			\end{itemize}
			\end{lemma}
			\begin{proof}
			The inequality sign in the derivation of the funnel  inequality~\eqref{eq:funnel_inequality} only comes from the use of Theorem~\ref{thm:[Si,Sj]>[Si][Sj]-K} in obtaining~\eqref{eq:Prob(X^j,A)-conditional}.  
			By Theorem~\ref{thm:[Si,Sj]>[Si][Sj]-K}, 
			inequality~\eqref{eq:Prob(X^j,A)-conditional} is
			strict if and only if 
			there exist $i_1,i_2 \in \{1, \dots, L\}$ and a node $m \in {\cal M}$ which	 is influential to  both~${j}_{A_l}$ and to~${j}_{A_{\widetilde{l}}}$, where
			$i_1 \not=i_2$.
			Since no edges emanate from~${j}_{A_1}, \dots, {j}_{A_L}$, 
			and~${j}_p$, we have that $m \in {\cal M} \setminus \{j\}$. 
			
			Thus, the funnel  inequality in strict if and only if 
			there exists a node
			$m \in {\cal M} \setminus \{j\}$ in~${\cal N}^+$ which is influential to~${j}_{A_l}$ and to~${j}_{A_{\widetilde{l}}}$.
			This, however, is the case if and only if 
			there exists a node $m \in {\cal M} \setminus \{j\}$
			which is influential to~${j}$ in~${\cal N}^{A_l}$ and in~${\cal N}^{A_{\widetilde{l}}}$,
			i.e.,  if  
			$j$ is not a funnel node of~$A_l$ and~$A_{\widetilde{l}}$.
			\end{proof}

			We can use the funnel  equality to compute the combined influences 
			from~$A_l$ and~$p_j$: 
			\begin{lemma}
			\label{lem:funnel_node_equality}
			Consider the Bass/{\rm SI} model~\eqref{eqs:Bass-SI-models-ME}.
			Let $j \in \cal M$ and $A_l \subset {\cal M} \setminus \{ j \}$.
			Then
			\begin{equation}
				\label{eq:funnel_only2}
				[S^{A_l,p_j}_j]  =
				[S^{A_l}_j] \,
				[S^{p_j}_j],  
				\qquad	l \in \{1, \dots, L\},
				\qquad t \ge 0,
			\end{equation}
			where $[S^{A_l,p_j}_j]  :=[S_j](t;{\cal N}^{A_l,p_j})$.
			\end{lemma}
			\begin{proof}
			Let~$\widehat{\cal N}$ denote the network obtained from~${\cal N}^{A_l,p_j}$ by adding 
			a fictitious isolated note, denoted by~$M+1$. Let $\widehat{\cal M}:=\{1, \dots, M+1\}$, 
			$B_1:={\cal M} \setminus \{j\}$, and $B_2:=\{M+1\}$. 
			Then
			$\{B_1,B_2, \{j\} \}$ is a partition of~$\widehat{\cal M}$,
			and  $j$~is a vertex cut, hence a funnel node, of~$B_1$ and
			$B_2$ in~$\widehat{\cal N}$.
			
			Let~$\widehat X_j$ denote the state of~$j$ in~$\widehat{\cal N}$. 
			By the funnel equality~\eqref{eq:funnel_equality},   
			$$
			[ \widehat S_j]
			=
			[ \widehat S^{B_1}_j] \,
			[ \widehat S^{B_2}_j] \,
			[ \widehat S^{p_j}_j] .
			$$
			By construction,
			\begin{align*}
				[ \widehat S_j]
				=[S^{A_l,p_j}_j], 
				\qquad 
				[ \widehat S^{B_1}_j] = [ \widehat S^{{\cal M} \setminus \{j\}}_j]
				=[S^{A_l}_j], 
				\quad 
				%
				[ \widehat S^{B_2}_j] = [ \widehat S^{\{M+1\}}_j]  \equiv 1, 
				\qquad 
				[ \widehat S^{p_j}_j] 
				=[S^{p_j}_j],
			\end{align*}
			where~$[S^{U}_j]$ denote the state of~$j$ in network~${\cal N}^U$.
			Therefore,  $	[S^{A_l,p_j}_j]  =
			[S^{A_l}_j] \,
			[S^{p_j}_j]$.  
			\end{proof}
			
			\begin{proof}[Proof of Theorems~{\rm \ref{thm:funnel_node_inequality-frac}} and~{\rm \ref{thm:funnel_node_inequality-frac-B}}]
			These theorems follow from Theorem~\ref{thm:funnel_node_inequality} and
			Lemmas~\ref{lem:funnel_node_iequality} and~\ref{lem:funnel_node_equality}. 
			\end{proof}
			
			\begin{proof}[Proof of Theorem~{\rm \ref{thm:funnel-upper-bound-frac}}]	
			The left inequality follows from~\eqref{eq:funnel_strict_inequality-A-p}.
			To prove the upper bound, we use the notations from the proof of Theorem~\ref{thm:funnel_node_inequality}. 
			By relations~\eqref{eq:Prob(X^j,A)=Prob(X^j,tilde-A)}, \eqref{eq:Prob(X^j,A)}, and~\eqref{eq:Prob(X^j,A)-indifference},  
			$$
			[S_j] = [S_j^{p_j}] \, [S^{+}_{{j}_{A_1}, \dots, {j}_{A_L}}].
			$$
			Recall that node~$j$ in network~${\cal N}$ 
			is split into nodes $\{{j}_{A_1}, \dots, {j}_{A_L}, {j}_{p_j}\}$ 
			in network~${\cal N}^+$. Hence, 	
			the cycle~$C_n$ corresponds  to a path~$\Gamma^+_n$  in~${\cal N}^+$  between some~${j}_{A_l}$ and~${j}_{A_{\widetilde{l}}}$ that 
			has $K_n+1$~nodes (including ${j}_{A_l}$ and ${j}_{A_{\widetilde{l}}}$).
			Therefore, by Lemma~\ref{lem:[S_i,j]-[S_i][S_j]-N-paths},
			$$
			[S^{+}_{{j}_{A_1}, \dots, {j}_{A_L}}]- \prod_{l=1}^L [S^{+}_{{j}_{A_l}}] 
			<
			\sum_{n=1}^{N_j} E(t;K_n+1), \qquad t>0.
			$$
			Multiplying this inequality by~$[S_j^{p_j}]$ and using the fact that $[S^{+}_{{j}_{A_l}}] = [S^{A_l}_{j}]$, see~\eqref{eq:Prob(X^j,A)=Prob(X^j,tilde-A)}, we obtain
			$$
			[S_j]  - [S_j^{p_j}] \prod_{l=1}^L [S_{j}^{A_l}] 
			<
			[S_j^{p_j}] \, \sum_{n=1}^{N_j} E(t;K_n+1), \qquad t>0.
			$$ 
			Since
			$[S^{A_l,p_j}_j]  =
			[S^{A_l}_j] \,
			[S^{p_j}_j]$, see~\eqref{eq:funnel_only2}, the inequality~\eqref{eq:funnel-upper-bound-1cycle} follows.
			\end{proof}

		\section{Final remarks}
		   \label{sec:Final}
		
		  In this study, we showed that 
		   $[S_{\cup_{l=1}^L\Omega_l}] - \prod_{l=1}^L [S_{\Omega_l}] \ge 0$, we found the necessary and sufficient condition for this inequality to be strict, and  obtained an upper bound for this difference.  We then used these results to derive the funnel theorems. 
		 These results enhance the arsenal of analytic tools for the Bass and SI models on networks.

	     While all of these results are new for the Bass model,
		some related results have appeared in the theory for epidemiological models.
	Thus,  for example, Cator and Van Mieghem~\cite{cator2014nodal} proved that $[S_{i,j}] \ge [S_i][S_j]$
		in epidemiological models where infected individuals 
		 can become susceptible again (SIS model) or recover  (SIR model).
		To the best of our knowledge, none of the theoretical studies  of  epidemiological models
		obtained the  necessary and sufficient condition under which this inequality is strict. Indeed, the role played by {\em influential nodes} is one of the methodological contribution of our study. In addition, to our knowledge, the upper bound for 
		$[S_{\cup_{l=1}^L\Omega_l}] - \prod_{l=1}^L [S_{\Omega_l}]$ 
		(Theorem~\ref{thm:[S_Omega]-prod[S_i]-N-paths}) was not obtained for 
		epidemiological models. 
				In~\cite{Kiss-15}, Kiss et al.\ derived 
		the funnel equality~\eqref{eq:funnel_equality} for the SIR model,
		for nodes that are vertex cuts. Our funnel theorems are more general
		in two aspects. First, we show that an equality holds not only 
		when the node is a vertex cut, but also when the node is a funnel node
		which is not a vertex cut. Second,  when the node is not 
		a funnel node, we obtain lower and upper bounds for 
		the funnel inequality.
		Finally, we note that the relation between the sign and magnitude of $[S_{\cup_{l=1}^L\Omega_l}] - \prod_{l=1}^L [S_{\Omega_l}]$ and of $[S_{{j}}] -\frac{\prod_{l=1}^L [S^{A_l, p_j}_j]}{([S^{p_j}_j])^{L-1}}$ (the funnel theorem)  was not noted in previous studies. 
		
	  In Sections~\ref{sec:applications-SiSj} and~\ref{sec:applications-funnel}  we showed some 
	  applications of these theoretical results to the BaSs and SI models on networks. 
	  We believe that there would be many more applications along the way. 
	  Moreover, one should be able to extend these results to other spreading models in epidemiology (SIR, SIS,\dots~\cite{Epidemics-on-Network-17})
	  to the Bass-SIR model~\cite{Bass-SIR-model-16}), as well as to 
	  spreading models on hypernetworks~\cite{Hyper-24}.

%
%

\bibliographystyle{plain}
\bibliography{../diffusion}

\appendix

	\section{Proof of~\eqref{eq:Prob(X^j,A)}}
\label{app:Prob(X^j,A)}

Let us fix $t>0$ and $N \in \mathbb{N}$. Let $\Delta t  = \frac{t}{N}$,
$t^N := N \Delta t$, 
and $X_{j}^n:=X_{j}(t^n)$. 
As $N \to \infty$, $\Delta t \to 0$ and $t^N \equiv t$.
Then we need to prove that 
\begin{equation}
	\label{eq:Prob(X^j,A)-discrete}
	\lim_{N \to \infty} [S_{{j}}](t^N;\Delta t) =\lim_{N \to \infty}
	[S^{+}_{{j}_{A_1}, \dots, {j}_{A_L},{j}_p}](t^N;\Delta t) .
\end{equation}
To do this, we introduce the following implementation of
the Bass/{\rm SI}  model~\eqref{eqs:Bass-SI-models-ME}:


	\begin{algorithmic}
		\State Choose~$\Delta t>0$
		\State  for  $j=1,\ldots,M$
		\State \qquad sample $\omega_{j}^{0}\sim U(0,1)$
		\State \qquad  { if} $0 \le \omega_{j}^{0} \leq I_j^0$ { then} ${X}_{j}^{0} := 1$ { else} ${X}_{j}^{0} := 0$ 
		\State  end 
		\State  for  $n=1,2,\ldots$
			\State \qquad  for  $j=1,\ldots,M$ 
			\State \qquad\qquad { if} ${X}_{j}^{n-1} = 1$ { then} ${X}_{j}^{n} := 1$
			\State \qquad\qquad { if} ${X}_{j}^{n-1} = 0$ { then}
			\State \qquad\qquad\qquad sample $\omega_{j}^{n}\sim U(0,1)$
			\State \qquad\qquad\qquad  { if} $0 \le \omega_{j}^{n} \leq \Big( p_j + \sum_{k\in \cal M}q_{k,j}  {X}_k^{n-1} \Big)\Delta t$ { then} ${X}_{j}^{n} := 1$ 
			\State \qquad\qquad\qquad { else} ${X}_{j}^{n} := 0$ 
			\State \qquad  end 
			\State  end 
		\end{algorithmic}
	\hrule
	\mbox{} \\
	
	Let us denote the outcome of this implementation by		
	$$
	\widetilde{X}_{k}^{N}
	:=X_{k}(t^{N};\{\vomega^{n}\}_{n=0}^\infty,\Delta t), 
	\qquad  k \in {\cal M}, \quad  N = 0,1, \dots
	$$
	where $\vomega^n := \{\omega_k^n\}_{k \in {\cal M}}$. 
	Let us also denote 
	$$
	\vomega_{-j}^n := \{\omega_k^n\}_{k \in {\cal M} \setminus \{j\}},
	\quad 
	\vomega^{+, n} = \{\vomega_{-j}^n, \omega_{{j}_{A_1}}^n,\dots, \omega_{{j}_{A_L}}^n,\omega_{{j}_{p}}^n\},
	\quad {\cal M}^+ := \big({\cal M} \setminus j \big) \cup \{{j}_{A_1}, \dots, {j}_{A_L},{j}_{p} \}.
	$$		
	The implementation  of
	the Bass/{\rm SI}  model~\eqref{eqs:Bass-SI-models-ME} on~${\cal N}^+$ is denoted by\,\footnote{The $L+2$ realizations $\omega_j^n, \omega_{{j}_{A_1}}^n,\dots, \omega_{{j}_{A_L}}^n,\omega_{{j}_{p}}^n$ are independent. }
	$$
	\widetilde{X}_{k}^{+, N} 
	:=X_{k}^{+}(t^{N};\{\vomega^{+, n}\}_{n=0}^\infty,\Delta t), \qquad  k \in {\cal M}^+, \quad  N = 0,1, \dots
	$$

	Since there are no edges that emanate from the nodes~$j, {j}_{A_1}, \cdots, {j}_{A_L}, {j}_p$,
	the sub-realizations $\{\vomega_{-j}^n\}_{n=0}^{\infty}$ completely determine $\{\widetilde{X}_{k}^N\}$ and  $\{\widetilde{X}_{k}^{+, N}\}$ for all~$	k \in {\cal M} \setminus \{j\}$ and~$N \in \mathbb{N}$.
	Hence, if we use the same $\{\vomega_{-j}^n\}_{n=0}^{\infty}$
	and~$\Delta t$ for both networks, then
	\begin{equation}
		\label{eq:tildeX_k(t^n)=X_k(t^N}
		\widetilde{X}_{k}^{N}
		\equiv
		\widetilde{X}_{k}^{+, N}, \qquad 
		k \in {\cal M} \setminus \{j\}, \quad N = 0,1, \dots 
	\end{equation}
	
	To compute the left-hand side of~\eqref{eq:Prob(X^j,A)-discrete},
	we first note that  
	$$
	\widetilde{X}_j^N=0 \quad \iff 
	\widetilde{X}_j^n=0 , \qquad n=0, \dots, N.
	$$
	Hence, 
	$$
	\widetilde{X}_j^N=0 \quad \iff \quad I_j^0<{\omega}_j^0 \le 1 ~~ \text{and} ~~ 
	{\omega}_j^n \ge  \Big(p_j+\sum\limits_{k \in {\cal M} \setminus \{j\}}  q_{k,j} \widetilde{X}_{k}(t^{n-1})\Big) \Delta t, \quad n=1, \dots, N.
	$$
	Therefore, 
	$$
	\Big[S_j  \mid  \{\vomega_{-j}^n\}_{n=1}^{N} \Big](t^N; \Delta t)
	= [S_j^0] \, \prod_{n=1}^{N} H_j^n   
	\qquad 
	H_j^n:=	
	1-\Big(p_j+\sum_{k \in {\cal M} \setminus \{j\}}  q_{k,j} 
	\widetilde{X}_{k}(t^{n-1})
	\Big) \Delta t,
	$$
	where $[S_j^0] = 1-I_j^0$.
	Hence, 
	\begin{align}
		\label{eq:P(X^{A_1, A_2,p}_j(t)=0)}
		&	
		[S_{{j}}](t^N; \Delta t)
		= 
		[S_j^0] 	\int_{[0, 1]^{(M-1)\times N}} \bigg( \prod_{n=1}^{N}  H_j^n\left(\{\vomega_{-j}^n\}_{n=1}^{N},\Delta t \right) \bigg) \,
		d\vomega_{-j}^{1}\cdots d\vomega_{-j}^{N}.
	\end{align}
	Similarly, to compute the right-hand side
	of~\eqref{eq:Prob(X^j,A)-discrete} , we note that
	$
	\widetilde{X}^{+, N}_{{j}_p}
	=\widetilde{X}^{+, N}_{{j}_{A_1}}
	= \cdots =\widetilde{X}^{+, N}_{{j}_{A_L}}
	=0
	$
	if and only if $\widetilde{X}^{+, 0}_{{j}_p}
	=\widetilde{X}^{+, 0}_{{j}_{A_1}}
	= \cdots =\widetilde{X}^{+, 0}_{{j}_{A_L}}=0$, and for $n=1, \dots, N$, 
	$$
	{\omega}_{{j}_p}^n \ge p_j \Delta t, \qquad 
	{\omega}_{{j}_{A_l}}^n \ge \Big(\sum\limits_{k \in A_l}  q_{k,j} \widetilde{X}^{+, n-1}_k \Big) \Delta t,
	\quad l \in \{1, \dots, L\}.
	$$
	Since $\widetilde{X}^{+, 0}_{{j}_{A_1}}  	= \cdots = \widetilde{X}^{+, 0}_{{j}_{A_L}} \equiv 0$, then $\mathbb{P}(\widetilde{X}^{+, 0}_{{j}_p}
	=\widetilde{X}^{+, 0}_{{j}_{A_1}}
	= \cdots = \widetilde{X}^{+, 0}_{{j}_{A_L}}=0) =  \mathbb{P}(\widetilde{X}^{+, 0}_{{j}_p}=0) =  [S_j^0]$.
	Therefore,
	\begin{equation}
		\label{eq:P(tildeX^{A,p}_j(t)=0)}
		[S^{+}_{{j}_{A_1}, \dots, {j}_{A_L},{j}_p}](t^N;\Delta t)	= 
		[S_j^0] \int_{[0, 1]^{(M-1)\times N}}
		\bigg( \prod_{n=1}^{N} H_j^{n, +}\left(\{\vomega_{-j}^n\}_{n=1}^{N}, \Delta t \right) \bigg)
		d\vomega_{-j}^{1}\cdots d\vomega_{-j}^{N},
	\end{equation}
	where
	$$
	H_j^{n, +}:= (1-p_j\Delta t)
	\prod_{l=1}^L \Big(1-  \Delta t \sum_{k \in A_l}  q_{k,j} \widetilde{X}^{+, n-1}_k \Big).
	$$
	
	To finish the proof of~\eqref{eq:Prob(X^j,A)-discrete}, 
	we now show that  as $N \to \infty$, the integrand 
	$\prod_{n=1}^{N} H_j^n$ of~\eqref{eq:P(X^{A_1, A_2,p}_j(t)=0)} approaches, uniformly in~$\{\vomega_{-j}^n\}_{n=1}^{N}$, the integrand $\prod_{n=1}^{N}	
	H_j^{n, +}$ of~\eqref{eq:P(tildeX^{A,p}_j(t)=0)}.\,\footnote{\rm Note that if we would have defined the adoption probability in~\eqref{eq:general_model} as $1-e^{-\lambda_j \Delta t}$ instead of
		$\lambda_j \Delta t$,  
		then
		$H_j^{n}$ and $H_j^{n, +}$ would have been identical. 
	} 
	Indeed,  
	by~\eqref{eq:tildeX_k(t^n)=X_k(t^N}, 
	\begin{equation*}
		\begin{aligned}
			H_j^{n, +}&
			=
			\left(1 - p_j\Delta t\right)
			\prod_{l=1}^L   \Big(1-\Delta t \sum_{k \in A_l}  q_{k,j} \widetilde{X}_k^{n-1}  \Big)
			\\			&=
			1-\Big(p_j+\sum_{k \in {\cal M} \setminus \{j\}}  q_{k,j} \widetilde{X}_k^{n-1} \Big) \Delta t+ D_j^n (\Delta t)^2,
		\end{aligned}
	\end{equation*}
	where 
	$$
	D_j^n := p_j \Big(\sum_{k \in {\cal M} \setminus \{j\}}  q_{k,j} \widetilde{X}_k^{n-1} \Big)  +
	(1- p_j \Delta t) 
	\prod_{l=1}^L \Big(\sum_{k \in A_l}  q_{k,j} \widetilde{X}_k^{n-1} \Big)
	.
	$$		 
	Hence,  
	$$
	H_j^{n, +}  = H_j^n + D_j^n (\Delta t)^2 = 	H_j^n 	\bigg(1+ \frac{D_j^n (\Delta t)^2}{H_j^n} \bigg),
	$$		  
		and so
		\begin{eqnarray}
			\label{eq:three-products}
			&&
			\prod_{n=1}^{N}	
			H_j^{n, +}
			= \prod_{n=1}^{N} 
			H_j^n
			\prod_{n=1}^{N} \bigg(1+ \frac{D_j^n(\Delta t)^2}
			{H_j^n}
			\bigg).
		\end{eqnarray}
		Thus, by~\eqref{eq:P(X^{A_1, A_2,p}_j(t)=0)}--\eqref{eq:three-products}, to finish the proof of~\eqref{eq:Prob(X^j,A)-discrete}, we need to show that
		\begin{equation}
			\label{eq:prod->1-uniform}
			\lim_{\Delta t \to 0} \prod_{n=1}^{N} \bigg(1+ \frac{D_j^n (\Delta t)^2}
			{H_j^n} \bigg)
			= 1,
		\end{equation}
		uniformly in $\{\vomega_{-j}^n\}_{n=1}^{N}$.

		The three sums that appear in~$D_j^n$ and in~$H_j^n$ are uniformly bounded:
		$$
		0 \le 
		\sum_{k \in A_l}  q_{k,j} \widetilde{X}_k^{n-1} 
		\le 
		\sum_{k \in {\cal M} \setminus \{j\}}  q_{k,j} \widetilde{X}_k^{n-1} 
		\le \sum_{k \in \cal M}  q_{k,j}  =q_j. 
		$$
		Hence,
		$$
		0 \le D_j^n \le q_j\left(p_j +q_j \right) ,
		\quad
		1-(p_j+q_j ) \Delta t \le  H_j^n \le 1, 
		$$	
		and so for $0<\Delta t \ll 1$,
		$$
		0 \le  \frac{D_j^n}{H_j^n} \le  \frac{q_j(p_j +q_j)}{1-(p_j+q_j ) \Delta t} \le 2 q_j\left(p_j +q_j \right). 
		$$
		%
		Therefore, 
		$$
		1 \le 	\prod_{n=1}^{N} \bigg(1+ \frac{D_j^n (\Delta t)^2}
		{H_j^n} \bigg)\le
		\prod_{n=1}^{N} \Big(1+2q_j \left(p_j +q_j \right) (\Delta t)^2 \Big). 
		$$
		Since $N = \frac1{\Delta t}$, 
		the right-hand side approaches~1 as $\Delta t \to 0$,  uniformly in $\{\vomega_{-j}^n\}_{n=1}^{N}$. Hence, we proved~\eqref{eq:prod->1-uniform}.

\end{document}